\newif\ifpdflatex    
\pdflatextrue           

\documentclass[fleqn,usenatbib,useAMS,twocolumn,numberedappendix]{aastex}
\usepackage{amsmath,esint}
\usepackage{url,aasmacros}
\usepackage{natbib}
\usepackage{soul, CJK}
\usepackage{multirow}
\usepackage{tablefootnote}
\usepackage{appendix}
\usepackage{bm}
\usepackage{xspace}
\usepackage{color}
\usepackage{enumitem}
\usepackage{booktabs}

\usepackage{graphicx}	
\usepackage{amsmath}	
\usepackage{amssymb}	
\usepackage{bm}		
\usepackage{url}
\usepackage{amsbsy}
\usepackage{xspace}
\usepackage{hyperref}
\usepackage{longtable}
\usepackage[colorinlistoftodos]{todonotes}
\usepackage[flushleft]{threeparttable}

\usepackage[T1]{fontenc}
\usepackage{ae,aecompl}

\def\lesssim{\mathrel{\hbox{\rlap{\hbox{\lower5pt\hbox{$\sim$}}}\hbox{$<$}}}}
\def\gtrsim{\mathrel{\hbox{\rlap{\hbox{\lower5pt\hbox{$\sim$}}}\hbox{$>$}}}}

\newcommand{\angstrom}{\textup{\AA}\xspace}

\usepackage{upgreek}                                                  
\usepackage{xspace}                                                       
\newcommand{\um}{$\upmu$m\xspace}            
 
\def\apjl{ApJ}%
\def\apjs{ApJS}%
\def\aap{A\&A}%
\shorttitle{WNTR23bzdiq}
\shortauthors{Karambelkar et al.}

\begin{document}
\title{The slow brightening of WNTR23bzdiq / WTP19aalzlk : Possible onset of common-envelope evolution in an asymptotic giant branch star?}

\author[0000-0003-2758-159X]{Viraj R. Karambelkar}
\email{viraj@astro.caltech.edu}
\affiliation{Cahill Center for Astrophysics, California Institute of Technology, Pasadena, CA 91125, USA}

\author[0000-0002-5619-4938]{Mansi M. Kasliwal}
\affiliation{Cahill Center for Astrophysics, California Institute of Technology, Pasadena, CA 91125, USA}

\author[0000-0002-8989-0542]{Kishalay De}
\affiliation{Columbia University, 538 West 120th Street 704, MC 5255, New York, NY 10027}
\affiliation{Center for Computational Astrophysics, Flatiron Research Institute, 162, 5th Ave, New York, NY 10010}

\author[0000-0002-7197-9004]{Danielle Frostig}
\affiliation{Center for Astrophysics | Harvard \& Smithsonian, 60 Garden Street, Cambridge, MA 02138, USA}

\author[0000-0003-2434-0387]{Robert Stein}
\affiliation{Joint Space-Science Institute, University of Maryland, College Park, MD 20742, USA}
\affiliation{Department of Astronomy, University of Maryland, College Park, MD 20742, USA}
\affiliation{Astrophysics Science Division, NASA Goddard Space Flight Center, Mail Code 661, Greenbelt, MD 20771, USA}

\author[0000-0001-6331-112X]{Geoffrey Mo}
\affiliation{MIT-Kavli Institute for Astrophysics and Space Research, 77 Massachusetts Ave., Cambridge, MA 02139, USA}

\author[0000-0002-4585-9981]{Nathan P. Lourie}
\affiliation{MIT-Kavli Institute for Astrophysics and Space Research, 77 Massachusetts Ave., Cambridge, MA 02139, USA}

\author[0000-0003-3769-9559]{Robert A. Simcoe}
\affiliation{MIT-Kavli Institute for Astrophysics and Space Research, 77 Massachusetts Ave., Cambridge, MA 02139, USA}

\author[0000-0003-0901-1606]{Nadejda Blagorodnova}
\affiliation{Institut de Ciències del Cosmos (ICCUB), Universitat de Barcelona (UB), c. Martí i Franquès, 1, 08028 Barcelona, Spain}
\affiliation{Departament de Física Quàntica i Astrofísica (FQA), Universitat de Barcelona (UB), c. Martí i Franquès, 1, 08028 Barcelona, Spain}
\affiliation{Institut d’Estudis Espacials de Catalunya (IEEC), c/ Esteve Terradas,1, Edifici RDIT, Despatx 212, Campus del Baix Llobregat UPC - Parc Mediterrani de la Tecnologia, 08860 Castelldefels, Spain}

\begin{abstract}
We present WNTR23bzdiq/WTP19aalzlk, a slow eruption of an early-asymptotic giant branch (AGB) star in M\,31 identified by the Wide-field Infrared Transient Explorer (WINTER) near-infrared and the NEOWISE mid-infrared surveyors. This source brightened gradually over seven years: a 0.5\,mag optical rise (2018-2021), a 1-mag optical outburst lasting $\sim$1000 days (2021- 2023), and another 1\,mag optical rebrightening in 2024. This was accompanied by a steady mid-IR brightening of 1\,mag over ten years in NEOWISE data. Archival optical data show only erratic, small amplitude ($<0.3$\,mag) brightness variations from 2003 to 2015, revealing a progenitor star with T$_{\rm{eff}} \approx 3500$\,K and L\,$\approx1.6\times10^{4}$\,L$_{\odot}$ -- consistent with a 7$\pm$2\,M$_{\odot}$ star in its early-AGB phase. During the eruption, the luminosity rose to $\approx5\times10^{4}$\,L$_{\odot}$ with slow photospheric expansion ($\approx5$\,km\,s$^{-1}$) and constant temperatures ($\approx3600$\,K) inferred from the spectral energy distribution. Optical and NIR spectra of the eruption resemble late M-type stars, with a mixed-temperature behavior -- transitioning from M1 in the optical to M7/M8 in the NIR. These properties of WNTR23bzdiq resemble those of stellar merger transients, particularly the giant star merger OGLE-2002-BLG-360, but on longer timescales. As such, WNTR23bzdiq potentially marks the onset of common-envelope evolution (CEE) in a binary with an AGB primary, and is possibly a member of the emerging population of infrared transients from CEE in giant stars. Continued multiwavelength monitoring, particularly mid-IR observations with \emph{JWST} to quantify dust production, will shed further light on WNTR23bzdiq. \\ \\ 
\end{abstract}

\section{Introduction}
Common envelope evolution (CEE) is an important evolutionary phase in the lives of binary stars \citep{Paczynski1976IAUS, Webbink84}. This phase is triggered when mass transfer in the binary system becomes dynamically unstable, causing the primary star to engulf the secondary star in a shared common envelope. Dynamical friction then triggers an inspiral of the binary system, which can result either in a stellar merger, or an envelope ejection that leaves behind a stable binary in a closer orbit \citep{Ivanova2013araa}. Thus, CEE can produce the rapid orbital shrinkage required to explain the formation of double compact objects in very close orbits that can merge by radiating gravitational waves within the Hubble time \citep{Dominik12, Ivanova2013araa, Postnov2014, VignaGomez2020, Marchant2021}. Despite its vital role in the formation pathways of gravitational wave sources, CEE remains poorly understood due to a lack of direct observations of this short-lived phase \citep{Ivanova2013araa}. 

The final stage of CEE -- merger or envelope ejection -- powers luminous stellar eruptions. A class of transients called Luminous Red Novae (LRNe) has been associated with binary systems undergoing CEE \citep{Pastorello2019a, Blagorodnova2021, Tylenda11, Kasliwal2011, Munari2002AA, Kulkarni07}. These events offer an important probe into understanding the physical processes involved in CEE. Most LRNe have been identified by optical time-domain surveys based on a sharp increase in the system's luminosity lasting tens to a few hundred days \citep{Karambelkar2023, Pastorello2021a, Pastorello2021b, Pastorello2022, Cai2022}. These eruptions typically exhibit multi-peaked lightcurves, a rapid transition from blue to red colors due to dust formation, and a spectroscopic transition to M-type spectra with strong molecular absorption features \citep{Munari2002AA, Tylenda11, Kochanek14_mergers, Pastorello2019a, Karambelkar2023}. While LRNe span a wide range of progenitor masses, the evolutionary stages are largely limited to stars on the main-sequence or stars crossing the Hertzsprung gap \citep{Blagorodnova17,MacLeod17,Blagorodnova2021, MacLeod2022}. Of the more than two dozen LRNe discovered to date, only one has a giant star progenitor \citep{Tylenda13}.

The paucity of CEE transients from giant stars  has been attributed to their markedly different appearance compared to typical LRNe \citep{MacLeod2022}. Owing to their extended envelopes and longer dynamical timescales, giant star merger transients are expected to evolve much more slowly than LRNe. Furthermore, these stars are predicted to lose a lot more mass upon the initiation of CEE and are expected to be more enshrouded in dust than their lesser evolved counterparts \citep{MacLeod2022}. Consequently, giant star mergers are predicted to manifest as slow-evolving, long-lived, primarily infrared transients -- significantly different from optical LRNe \citep{MacLeod2022, Pejcha16b, Metzger2017}. 

The Galactic transient OGLE-2002-BLG-360 is the only CE transient with a confirmed red giant progenitor \citep{Tylenda13}. This eruption, originating from a K-type giant star, lasted for over two thousand days, exhibited multiple lightcurve peaks along with a spectral evolution to an M-type star at late times. While its overall morphology resembled LRNe, its duration was nearly an order of magnitude longer than typical optical LRNe. Infrared and sub-mm observations show that it is the dustiest among Galactic stellar mergers \citep{Steinmetz2025}, establishing it as the prototypical event for a slow, dusty transient from CEE in a binary system with a giant primary. Beyond OGLE-BLG, the historical long-duration eruption of CK\,Vul \citep{Kato2003, Tylenda13} and the population of IR transients discovered by the \emph{Spitzer Space Telescope} without any optical counterparts (termed SPRITES, \citealt{Kasliwal2017ApJ, Jencson2019_spirits}) have been attributed to mergers or envelope ejections involving evolved, giant stars. Nonetheless, the phase space of giant star mergers and envelope ejections remains largely unexplored, primarily due to a lack of infrared time-domain surveys. 

In this paper, we present WNTR23bzdiq/WTP19aalzlk -- a slow, ongoing infrared eruption of a giant star in  M\,31, whose properties align with the expectations for CEE involving a giant star. This source was independently identified by two infrared searches: the new Wide-field Infrared Transient Explorer (WINTER, \citealt{Lourie2020, Frostig2024}) in the near-IR (NIR), and a dedicated pipeline for detecting mid-IR transients using the NEOWISE survey \citep{De2023Nature, De2024_m31, Meisner2019, Mainzer2014ApJ}. Archival photometry indicates that the giant progenitor of WNTR23bzdiq was an early asymptotic giant branch (AGB) star, making it the most evolved star associated with a possible CE transient. 

This paper is organized as follows -- Section \ref{sec:observations} describes the observations of this source, Section \ref{sec:analysis} analyzes the photometric and spectroscopic properties, Section \ref{sec:discussion} discusses possible origin scenarios for this eruption, and Section \ref{sec:summary} concludes with a summary of our results and path forward.

\begin{figure*}[hbt]
    \centering
    \includegraphics[width=\textwidth,angle=0]{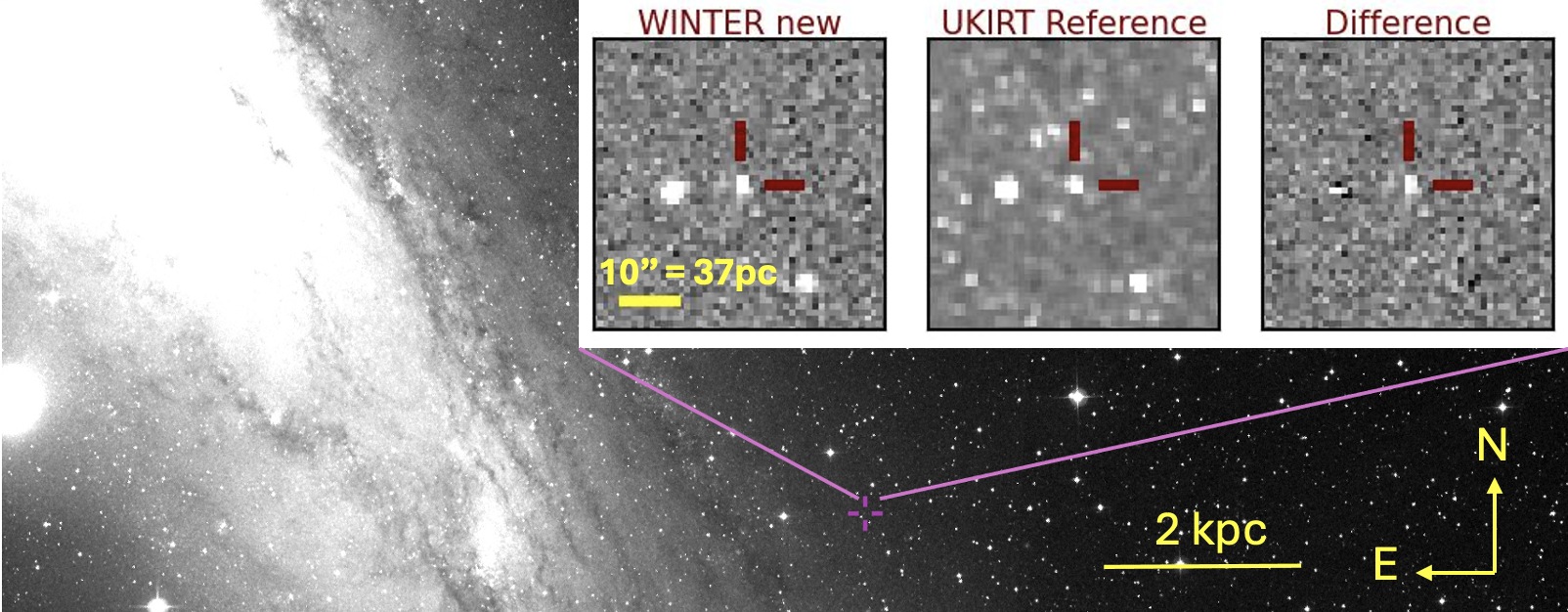}
    \caption{WNTR23bzdiq (pink cross) is located in the outskirts of M31, at an angular separation of  $\approx0.9$\,deg from its center corresponding to a distance of $\approx$12\,kpc. \emph{Inset:} A comparison of a new \emph{J-}band WINTER image from 2023, an archival \emph{J-}band UKIRT Hemisphere survey reference image from 2014, and the difference image of these two. The position of WNTR23bzdiq is marked by cross-hairs. The background image is taken from the Digitized Sky Survey II.}
    \label{fig:cutouts}
\end{figure*}

\section{Observations}
\label{sec:observations}
\subsection{Identification}
The WINTER surveyor is a 1.2\,sq\,deg. field-of-view InGaAs camera mounted on a 1\,m robotic telescope at Palomar observatory operating at the NIR \emph{Y}, \emph{J}, and \emph{Hs} bands \citep{Lourie2020, Frostig2024}. WINTER is conducting a Nearby Galaxies Survey that routinely images galaxies within 25\,Mpc in the \emph{J}-band to search for infrared transients. The source WNTR23bzdiq was identified as a candidate transient in the outskirts of M\,31 in the WINTER data stream on UT 2023-12-17 at $\alpha$=00:38:48.6, $\delta=$+40:46:08.2 by its automated transient detection pipeline. The full details of the WINTER transient detection pipeline will be outlined in a forthcoming paper (Karambelkar \& Stein et al., in prep). Briefly, the pipeline generates astrometrically and photometrically calibrated stacks of dithered observations and performs image subtraction relative to archival images from the UKIRT Hemisphere Survey (UHS, \citealt{ukirt_hemisphere}). Transient candidates identified from the difference images are broadcast to Skyportal/Fritz \citep{Coughlin2023} for human scanning and vetting of interesting sources. WNTR23bzdiq was flagged during scanning, and was coincident with a point source in the archival UHS image, suggesting that it was a star that had brightened since the UHS epoch. Figure \ref{fig:cutouts} shows cutouts of the new WINTER science image, UHS reference image, and difference image of the detection.

This source was independently identified as WTP19aalzlk -- a large-amplitude, long-duration mid-IR transient -- by the NEOWISE transient detection pipeline \citep{De2024_m31}, which searches for transients in unWISE stacks \citep{Meisner2019} of NEOWISE images. The source has been steadily brightening by $\approx1$\,mag in both W1 and W2 filters for the last ten years.  

\subsection{Photometry}
Since its discovery, WNTR23bzdiq was observed in the \emph{J}-band at several epochs as part of the regular WINTER survey. We performed aperture photometry at the position of the source in the stacked WINTER images to construct a \emph{J} band lightcurve. 

To supplement the WINTER \emph{J-}band observations, we compiled publicly available optical data from the Zwicky Transient Facility (ZTF, \citealt{Bellm2019}). This field was observed as part of the ZTF routine survey in the \emph{g,r,} and \emph{i} bands. Science-frame photometry up to UT 2024-06-29 was retrieved from the source catalogs in ZTF Data Release 22. Photometry for the later epochs was obtained through the ZTF forced photometry service \citep{Masci2019}. The difference fluxes from the forced photometry service were converted to the science fluxes adding flux of the nearest source in the reference image to the forced photometry. The ZTF fluxes were binned into 5-day bins to increase their signal-to-noise ratios.

Mid-infrared observations from the NEOWISE mission \citep{Mainzer2014ApJ} from 2014 to 2024 were obtained from the NEOWISE Single exposure source table in the final data release hosted on IRSA\footnote{https://irsa.ipac.caltech.edu/Missions/wise.html}. The NEOWISE fluxes from each pointing were binned together. 

\begin{figure*}[hbt]
    \centering
    \includegraphics[width=0.9\textwidth]{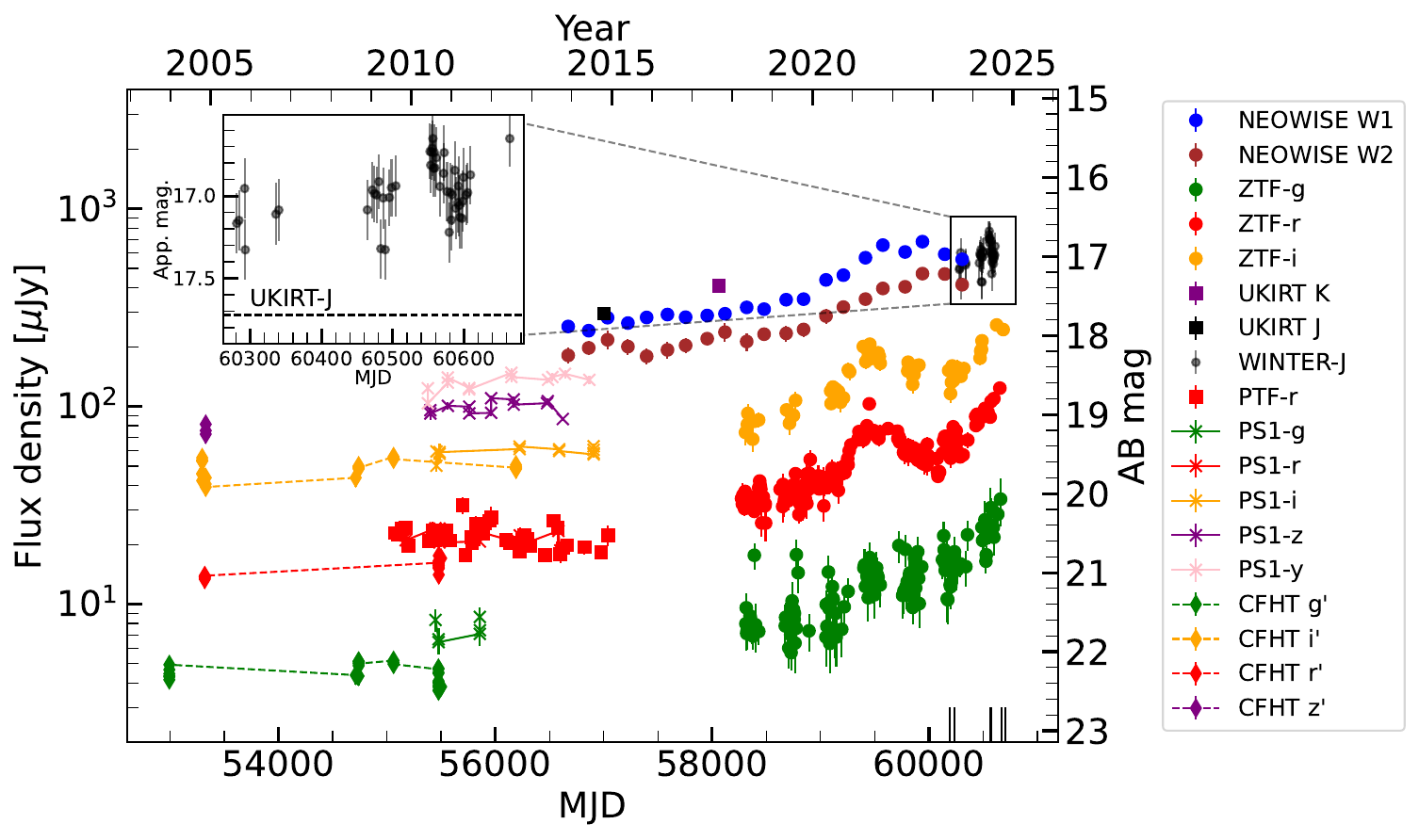}
    \caption{Multiband lightcurve of WNTR23bzdiq. The \emph{J-}band discovery lightcurve from WINTER is displayed in the inset plot, which shows a slow brightening from 2023 to 2024 relative to the 2014 UKIRT measurement. The full plot shows the optical to mid-IR photometry spanning 2003 to 2025. In the optical data available until 2015, the source only shows small amplitude ($<0.3$\,mag) erratic brightness variations. In 2018, it starts brightening smoothly, undergoes a 1000-day outburst from 2021 to 2023, and has continued rebrightening since. The NEOWISE mid-IR data shows a smooth brightening from 2016 to 2023, following which it decines slightly. Importantly, the mid-IR fluxes continued to increase even when the optical luminosity faded in 2023. The black vertical lines mark the epochs of the spectroscopic observations.}
    \label{fig:full_lc}
\end{figure*}

Archival optical observations were compiled from the Palomar Transient Factory (PTF, \citep{Law09}), the intermediate Palomar Transient Factory (iPTF \citealt{Cao16}), the PanSTARRS survey (PS1, \citealt{Chambers16, Magnier13, Flewelling2020}), and the Canada France Hawaii Telescope (CFHT). PTF/iPTF \emph{g} and \emph{r-}band lightcurves from 973 observations taken from 2010 to 2015 were downloaded from the point-source catalogs \citep{Ofek12, Laher14} hosted at IRSA\footnote{https://irsa.ipac.caltech.edu/Missions/ptf.html}. The PTF/iPTF observations were binned into 30-day bins to increase fidelity. PS1 \emph{grizy} fluxes for ten epochs spanning 2010 to 2015 were downloaded from the PS1 source catalogs hosted on MAST\footnote{https://catalogs.mast.stsci.edu/panstarrs/}. Even older observations from 2003 to 2012 were obtained from the CFHT MegaPrime data archive.\footnote{https://www.cadc-ccda.hia-iha.nrc-cnrc.gc.ca/en/cfht/} Reduced and calibrated \emph{g', r',i'} and \emph{z'-}band MegaPrime images were downloaded from the archive and photometry was performed at the location of WNTR23bzdiq to construct the lightcurves. Archival infrared observations were compiled from the 2MASS point source catalog (\citealt{Cutri2003}; one epoch each in \emph{JHKs} bands), the United Kingdom Infrared Telescope (UKIRT) M31 catalog (\citealt{Neugent2020}; one epoch each in \emph{J} and \emph{Ks} bands), the ALLWISE data release (\citealt{Cutri2014}; detections in W1 and W2 bands), and the Spitzer IRAC point source catalogs of M31 (\citealt{Khan2017}; one epoch, detections in channels 1, 2, and 3). Finally, archival \emph{Gaia} observations from 2016 confirm the membership of this source in M31 (see \citealt{Neugent2020}).

These optical and infrared lightcurves of WNTR23bzdiq are shown in Figure \ref{fig:full_lc}. Overall, we find good agreement between different photometry sources where they overlap, with small differences ($\sim 0.1$ mag) likely due to differences in filter profiles. Fig. \ref{fig:full_lc} suggests that this source did not show any significant brightness variations over a fifteen year period (2003 to 2018), following which it underwent an outburst lasting $\approx$1000 days, and has brightened again subsequently. 

\subsection{Spectroscopy}
\begin{figure*}[hbt]
    \centering
    \includegraphics[width=\textwidth]{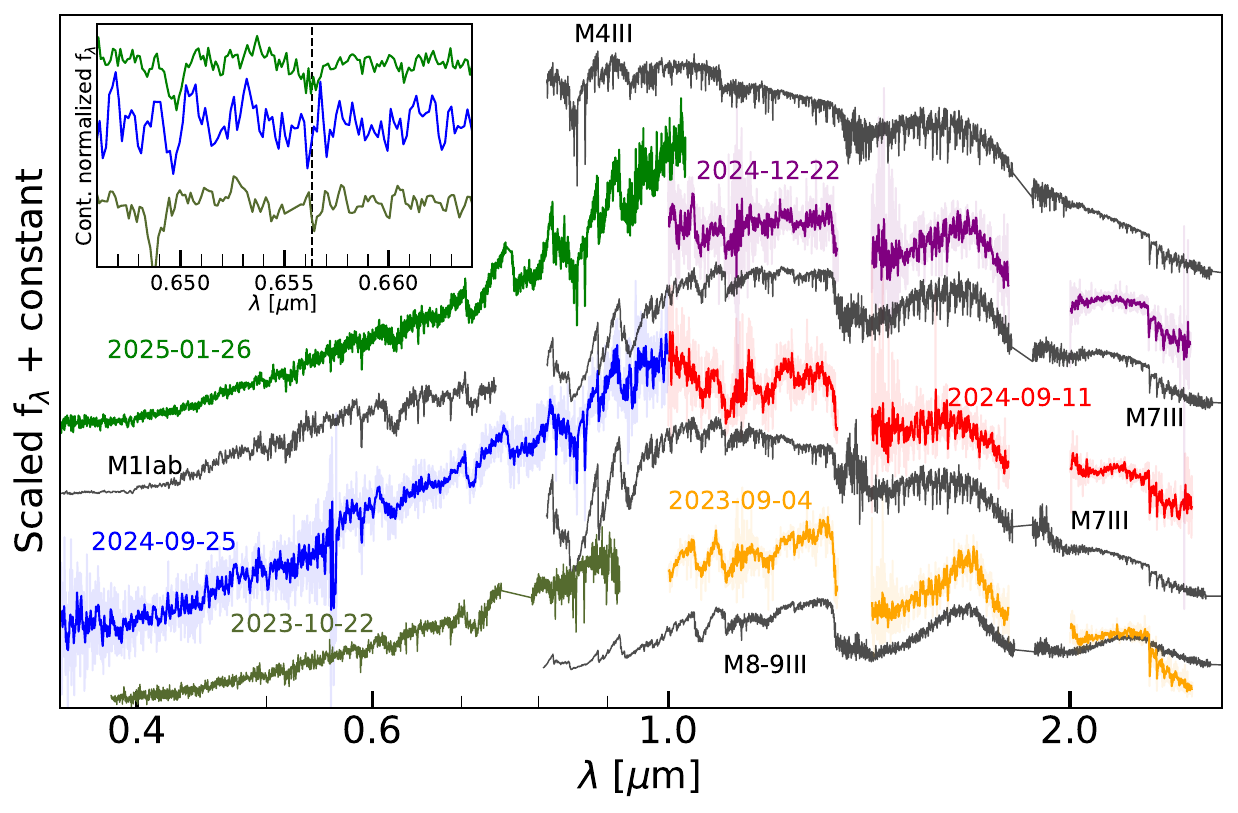}
    \caption{Optical-NIR spectra of WNTR23bzdiq (colored) compared to stellar spectral templates (black). The spectra of WNTR23bzdiq resemble those of late M-type giant stars. We find that the NIR spectral type varies from an M8-9 in our earliest spectrum that was taken at the end of the 2021 outburst (orange) to M7 in the next two spectra (red, purple) taken during the 2024 rebrightening. Additionally, the spectra at the two later epochs have mixed-spectral types, where the optical spectra match an M1 spectral type template while the NIR spectra match later M7 type templates. This behavior resembles the Galactic stellar merger V838\,Mon. \emph{Inset:} Continuum normalized optical spectra zoomed-in around the H$\alpha$ line, showing no strong H$\alpha$ emission.} 
    \label{fig:spectral_collage}
\end{figure*}

We obtained three optical spectra of this source, one with the Binospec spectrograph \citep{Fabricant2019} on the MMT telescope (PI: MacLeod) on 2023-10-22 one with the Double Beam Spectrograph (DBSP, \citealt{Oke82}, $R\approx1000$) on the 200-inch Hale telescope at Palomar Observatory on 2024-09-25, and one with the Low Resolution Imaging Spectrometer (LRIS, \citealt{Oke95}, R$\approx750$) on the Keck I telescope at Mauna Kea Observatory on 2025-01-26. Additionally, we obtained three near-infrared spectra, two with the Near-Infrared Echelle Spectrograph (NIRES, \citealt{Wilson2004}, R$\approx2500$) on the Keck II telescope on 2023-09-04 (NASA Program ID: 2023B-N258, PI: De) and 2024-09-11, and one with the Triplespec spectrograph (\citealt{Herter2008}, R$\approx2700$) on the 200-inch Hale telescope on 2024-12-22. The DBSP spectrum was reduced using the python package \texttt{dbsp\_drp} \citep{MandigoStoba2021}. The LRIS spectrum was reduced using the \texttt{idl} package \texttt{lpipe} \citep{Perley2020}. The NIRES and TripleSpec spectra were reduced using the \texttt{idl} package \texttt{spextool} \citep{Cushing2004} and corrected for telluric absorption using the package \texttt{xtellcor} \citep{Vacca2003}. These optical and NIR spectra are shown in in Figure \ref{fig:spectral_collage}. The black vertical lines in Figure \ref{fig:full_lc} mark the epochs of the spectroscopic observations. The spectra show strong absorption bands of oxygen-rich molecules and resemble late M-type stars. The locations of the molecular absorption bands are consistent with the redshift of M31, confirming the M31 membership of this source. 

\subsection{Distance and reddening}
For the analysis in this paper, we assume a distance to M\,31 of 0.77\,Mpc \citep{Ferrarese2000}. We correct for a Galactic line-of-sight extinction of $E_{B-V} = 0.09$\,mag \citep{Schlafly12}. We do not adopt any additional reddening from M\,31, as the source is located on the outskirts of the galaxy where the interstellar dust obscuration is expected to be low, and we do not find any evidence for additional reddening in our modeling.
\section{Analysis}
\label{sec:analysis}
\subsection{Progenitor modeling}
\label{sec:progenitor}
\begin{figure*}[hbt]
    \centering
    \includegraphics[width=0.5\textwidth]{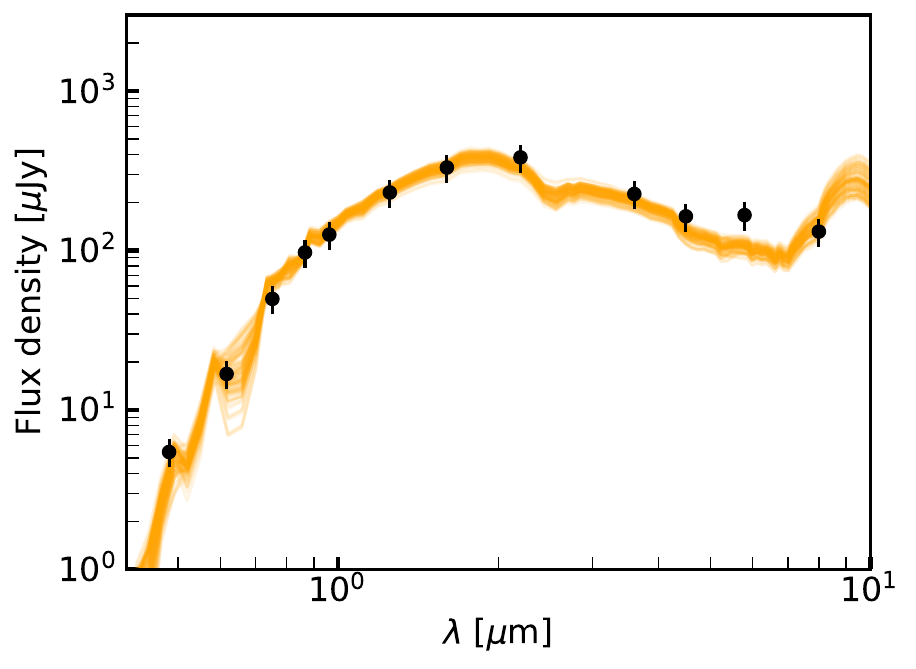}\includegraphics[width=0.5\textwidth]{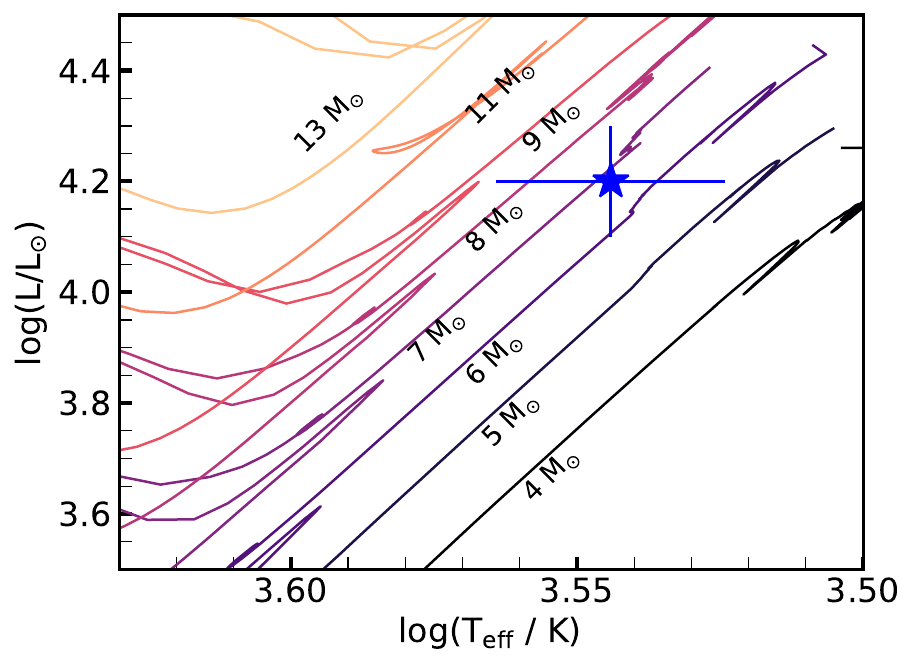}
    \caption{\emph{Left:} Spectral energy distribution of the progenitor star (black dots) together with the range of best-fit PHOENIX+DUSTY models (orange). Our modeling suggests the progenitor has T$_{\rm{eff}}\approx3500$\,K and a total luminosity log\_L$=4.2$\,L$_{\odot}$. \emph{Right}: Comparing these progenitor properties (blue star) to MIST stellar evolutionary tracks (colored lines) in the HR diagram suggests the progenitor is a 7$\pm$2\,M$_{\odot}$ star in its early AGB phase. }
    \label{fig:progenitor_sed_hr}
\end{figure*}
We use the archival photometry to model the nature of the progenitor star of WNTR23bzdiq. We construct the progenitor's spectral energy distribution (SED) using the PS1 \emph{grizy} band, the UKIRT \emph{J} and \emph{Ks} band, 2MASS \emph{H-}band, ALLWISE 3.4 and 4.6\um, and Spitzer IRAC 5.8 and 8.0\um fluxes. We first fit the SED using a combination of PHOENIX model photospheres \citep{Husser2013} and the radiative transfer code \texttt{DUSTY} \citep{Ivezic97, Ivezic99}. We downloaded solar metallicity PHOENIX stellar models available online\footnote{https://www.stsci.edu/hst/instrumentation/reference-data-for-calibration-and-tools/astronomical-catalogs/phoenix-models-available-in-synphot} and used them as inputs to \texttt{DUSTY}. \texttt{DUSTY} models the emergent radiation from the star through a dust shell, assuming spherical symmetry. We used the log\_g=0.0 PHOENIX models for our analysis. For the \texttt{DUSTY} models, we assumed silicate dust grains (motivated by the oxygen-rich molecular features in our spectra) with a standard MRN distribution for dust grain sizes, a wind-like $r^{-2}$ radial density profile for the dust shell, and a shell-thickness ratio of 2.0. First, we computed a grid of \texttt{DUSTY} models for stars with T$_{\rm{eff}}$ ranging from 2100 to 4500\,K with increments of 100\,K (limited by availability of the PHOENIX models), dust temperatures T$_{\rm{d}}$ ranging from 500 to 2000\,K with increments of 100\,K, and for fifteen values of the dust optical depth in the \emph{V-}band (0.55\,\um) $\tau_{\rm{V}}$ uniformly sampled in log space from $\tau_{\rm{V}}$ = 0.01 to 10. Then, we interpolated between this grid using a convolutional neural network (CNN) model. The CNN model was implemented using \texttt{pytorch} \citep{pytorch}, comprises two hidden dense layers (with 64 and 128 units respectively) with ReLU activation, and an output layer with 203 units that predicts the SED from $0.1-20$\,\um for three input parameters (T$_{\rm{eff}}$, T$_{\rm{star}}$, $\tau_{\rm{V}}$) described above. The model was trained using the Adam optimizer \citep{Kingma2014} with an initial learning rate of 0.001 for 100 epochs using the grid of 6000 models with a 80-20\% train-test split. We find that the CNN model is able to predict the \texttt{DUSTY} SEDs very well, with differences between the predicted and true SEDs smaller than 1\% in the wavelength range from 0.1 to 20\,\um, well below the errorbars on the observations. We used this model to derive posterior distributions of T$_{\rm{eff}}$, T$_{\rm{dust}}$, and $\tau_{\rm{V}}$ using Markov Chain Monte Carlo (MCMC) sampling implemented in the \texttt{python} package \texttt{emcee} \citep{Mackey-Foreman2013}. 

Our \texttt{DUSTY} modeling suggests that the progenitor is a star with T$_{\rm{eff}} = 3500 ^{+150}_{-150}$\,K, surrounded by a dust shell with temperature T$_{\rm{d}} = 1450^{+130}_{-100}$\,K and optical depth $\tau_{V} = 2.7_{-0.4}^{+0.4}$. Figure \ref{fig:progenitor_sed_hr} (left panel) shows the range of best-fit models to the progenitor SED. We note that for such optical depths, the optical and NIR data can place strong constraints on T$_{\rm{eff}}$, but mid-IR data at longer wavelengths (especially covering the 10\,\um silicate feature) are required for strong constraints on the dust parameters. As our SED coverage ends at 8\,\um, it is possible that our dust estimates are subject to inaccuracies or error-underestimations in the mid-IR Spitzer bands. For this reason, while we are confident about the effective temperature estimates, we caution that our dust estimates may not be accurate. The total integrated luminosity of the progenitor obtained from the \texttt{DUSTY} fitting is log(L/L$_{\odot}) = 4.20\pm0.02$.

To test for any systematics associated with our \texttt{DUSTY} modeling, we also fit the progenitor SED using the Grid of Red supergiant and Asymptotic Giant Branch ModelS (GRAMS; \citealt{Sargent2010,Sargent2011,Srinivasan2011}) and the approach described in \citet{Jencson2022}. This suite of radiative transfer models consists of a base grid of 1225 spectra of stars with constant mass-loss rates and spherically symmetric shells of silicate dust around them. The spectra are computed using PHOENIX model photospheres \citep{Kucinskas2005,Kucinskas2006} combined with the dust radiative transfer code \texttt{2-Dust} \citep{Ueta2003}. The published grid uses 1\,M$_{\odot}$ stars with effective temperatures T$_{\rm{eff}}$ between 2100 and 4700\,K, at a fixed subsolar metallicity log(Z/Z$_{\odot})=-0.5$ and a fixed surface gravity log\,$g = -0.5$. The circumstellar dust is characterized by the dust temperature at the inner radius of the dust shell $T_{in}$ and the optical depth at 1\um $\tau_{1}$, which can be converted to a dust mass-loss rate $\dot{M_{d}}$ assuming a wind speed. An additional parameter is the inner radius of the dust shell, R$_{\rm{in}}$. Similar to \citet{Jencson2022}, we find that our derived parameters are largely insensitive to this parameter, and we fix it to R$_{\rm{in}}=11.0$\,R$_{*}$. The best-fit model has T$_{\rm{eff}} = 3500$\,K $\tau_{1} = 0.46$ (corresponding to $\tau_{V}=1.9$ for silicate dust), T$_{\rm{in}} = 1000$\,K -- consistent with the values from the \texttt{DUSTY} modeling. The total luminosity of the progenitor from the GRAMS modeling is log(L/L$_{\odot}) = 4.20^{+0.06}_{-0.03}$.

In summary, our modeling suggests that the progenitor is a cold star with conservative estimates of T$_{\rm{eff}}=3500^{+200}_{-200}$ and luminosity log(L/L$_{\odot})=4.2\pm0.1$, moderately enshrouded in hot dust with temperature T$_{\rm{dust}}\approx1000-1500$\,K and $\tau_{V}\approx2-3$. The photospheric radius is calculated to be $350\pm50$\,R$_{\odot}$ while the inner radius of the dust shell is $\approx1800 \pm 400$\,R$_{\odot}$. We compare the luminosity and effective temperature to stellar evolutionary tracks from MESA Isochrones and Stellar Tracks (MIST; \citealt{Dotter2016, Choi2016})\footnote{downloaded from https://waps.cfa.harvard.edu/MIST/} in the right panel of Figure \ref{fig:progenitor_sed_hr}. Comparing the position of the star in the HR diagram to the stellar tracks suggests that the progenitor is a star with a zero-age main-sequence (ZAMS) mass of $7\pm2$\,M$_{\odot}$ in its early-AGB phase. 

\subsection{Lightcurve evolution}
The multiband lightcurve of WNTR23bzdiq is shown in Figure~\ref{fig:full_lc}. No substantial variations ($>0.3$\,mag) are seen in the available optical data from 2003 to 2018. Starting in 2018, the ZTF data reveal a gradual brightening --- an increase of a few tenths of a magnitude over three years. This was followed by a 1\,mag eruption beginning in 2021, which lasted for roughly 1000 days. The star faded towards the end of 2023, then rebrightened in 2024 and continues to brighten at the time of writing. The NEOWISE data show a smooth and steady increase in the mid-IR brightness from 2016 to 2023, followed by a decline in the next year. The mid-IR fading lags behind the optical, with the mid-IR fading beginning about a thousand days after the optical fading. No mid-IR data are available during the 2024 rebrightening.  

\begin{figure}
    \centering
    \includegraphics[width=0.5\textwidth]{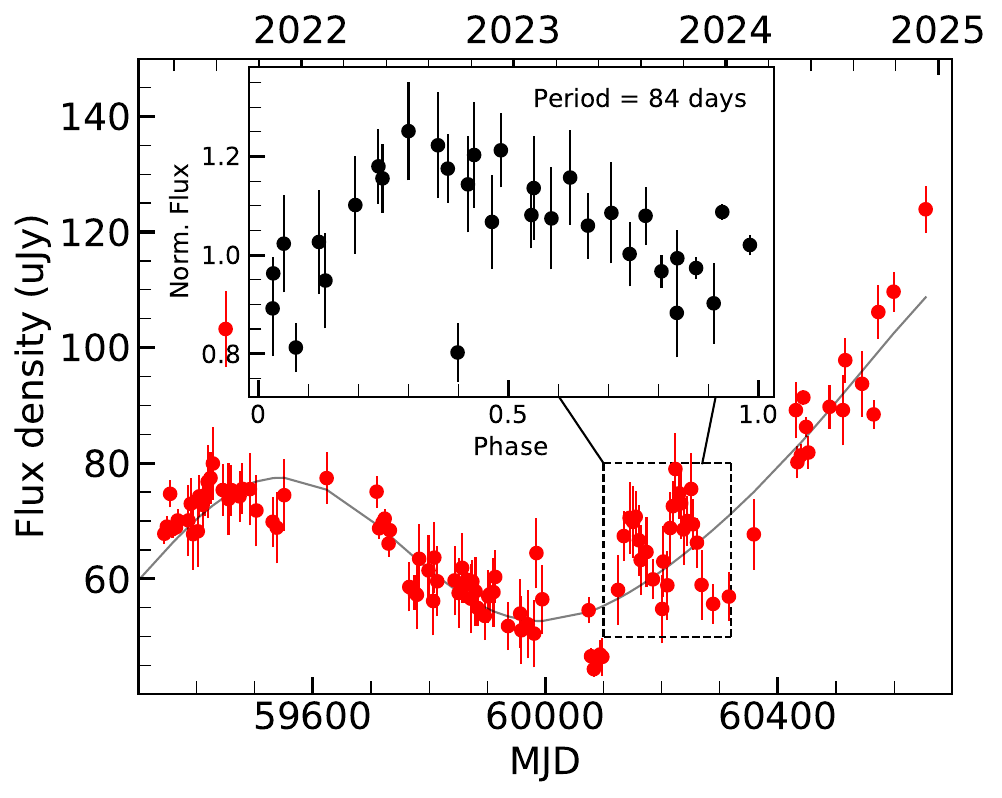}
    \caption{The \emph{r-}band lightcurve (red points) shows two successive bumps of $\approx 0.2$\,mag amplitude (dashed square) at the end of the first eruption in 2023. The inset shows a phase-folded lightcurve of these two bumps, where the flux has been normalized by a smooth gaussian process fit (black line) and phase folded at the derived period of 84 days. However, these two sharp bumps are unlikely to be caused by orbital motion of the binary, but could be due to accretion episodes on the secondary.}
    \label{fig:periodicity}
\end{figure}

We do not find any clear periodic signatures in the lightcurve. At the end of the first eruption in 2023, the \emph{r-}band lightcurve shows two sharp, consecutive bumps, each $\approx 0.2$\,mag in amplitude (Figure \ref{fig:periodicity}). A Lomb Scargle analysis of this segment yields a period of $\approx84$\,days. However, these two sharp bumps are unlikely to be caused by orbital motion of the binary, but could be due to accretion episodes on the secondary. Outside this, we do not detect any periodic signatures in the ZTF or the archival PTF data, although the larger uncertainties in the PTF and early ZTF data limit sensitivity to low-amplitude variations. 

We derived the evolution of the bolometric luminosity, effective temperature, stellar radius, and dust properties using the multi-band light curves, applying \texttt{DUSTY} modeling combined with MCMC sampling as described in Section \ref{sec:progenitor}. To build spectral energy distributions (SEDs) at several epochs between 2018 and 2025, we used Gaussian Process (GP) regression to interpolate the light curves and fill temporal gaps. The GP model, implemented with the \texttt{python} package \texttt{scikit-learn} \citep{scikit-learn}, employed a Radial Basis Function kernel with an added White kernel to account for observational noise. For epochs between 2018 and 2023, the SEDs include five bands (\emph{griW1W2}). We find that the \emph{W1} and \emph{W2} measurements are essential for accurately constraining the bolometric luminosity; omitting them results in underestimates by factors of up to two. In contrast, for 2024 and 2025, only optical (\emph{gri}) and NIR (\emph{J}) data are available. As a result, we cannot accurately estimate stellar parameters during this period. We report values by assuming the $r-W1$ and $r-W2$ colors remain similar to those in late 2023 -- an assumption supported by the minimal color variations observed in the previous lightcurve.

In 2018, the bolometric luminosity was $\approx20000$\,L$_{\odot}$, already brighter than the progenitor's 2015 luminosity of $\approx16000$\,L$_{\odot}$. This slow brightening continued through 2020, after which the luminosity increased more rapidly, reaching $\approx40000$\,L$_{\odot}$ by the end of 2022.  Notably, this luminosity increase began $\approx200$ days prior to the \emph{r-}band outburst and was driven primarily by the brightening in the \emph{W1} and \emph{W2} bands. The luminosity plateaued for the next five hundred days before declining slightly to $\approx35000$\,L$_{\odot}$ in late 2023. Based on the 2024 data, we infer a further increase to $\approx 50000$,L$_\odot$ by the end of that year; however, we caution that these estimates are uncertain due to the lack of mid-IR data during this time. 

The effective temperature in 2018 is estimated to be $\approx 3650$\,K, marginally (1$\sigma$) warmer than the progenitor. Over the subsequent years, the temperature remains nearly constant, with small fluctuations of $\approx 50$\,K that are consistent with the NIR spectroscopic evolution (Section \ref{sec:spec}). For instance, the temperature in 2024 is predicted to be slightly higher than in 2023, consistent with the warmer spectral types inferred from the 2024 NIR spectra.

The dust optical depth remains roughly constant between 2018 and mid-2022 at $\tau_V \approx 3.2$, slightly higher than that of the progenitor ($2.7 \pm 0.4$). In early 2023, $\tau_V$ increases to $\approx 3.6$ before declining thereafter. This evolution mirrors the optical light curve: the optical depth peaks when the mid-IR flux is at its maximum and the optical emission is fading. During this time (2018 to 2025), the dust shell temperature remains steady at $\approx 1200$,K, while the inner radius of the shell expands from $\approx 3000$\,R$\odot$ to $\approx 5000$\,R$\odot$. We estimate the dust mass from the \texttt{DUSTY} model parameters using the prescription from \citet{Lau2025}, finding an increase from $\sim 1 \times 10^{-7}$\,M$_{\odot}$ to $\sim 4 \times 10^{-7}$\,M$_{\odot}$ over this period. While the inferred dust evolution aligns well with the multi-wavelength photometric changes, we reiterate that these estimates are based on data at wavelengths $\lesssim 5$,$\mu$m, where the impact of dust is not most prominent. Observations at longer wavelengths are essential to robustly constrain the dust properties and their evolution in this system.

\subsection{Spectroscopic evolution}
\label{sec:spec}
Figure \ref{fig:spectral_collage} shows the optical and near-IR spectra of WNTR23bzdiq, along with the closest matching stellar spectral templates from the Jacoby-Hunter-Christian Atlas \citep{Jacoby1984} and the IRTF spectral library \citep{Rayner2009}. The spectra generally resemble late M-type giant stars; however, the spectral sub-types evolve between epochs. No strong emission lines are detected in any of the spectra.

Our earliest NIR spectrum, taken towards the end of the first outburst, shows strong absorption bands of TiO, H$_{2}$O, and CO. The \emph{H}-band bump shaped by water-vapor absorption on either sides resembles that of very late M8-M9 type stars. An optical spectrum from a month later also exhibits molecular absorption bands of TiO and atomic absorption such as the Ca H$\&$K, \ion{Na}{1}D doublet, Mg, and the Ca NIR triplet. This optical spectrum is more consistent with an M1-type star -- in contrast to the much colder NIR spectrum -- implying a ``mixed-temperature". 

The next set of spectra -- an optical and an NIR pair taken two weeks apart in September 2024, roughly a year into the second brightening -- show similar behavior. The NIR spectrum continues to show strong molecular absorption bands, though with a noticeable change in the appearance of the \emph{H}-band shape, now resembling an M7-type star. The optical spectrum remains consistent with an M1-type. Even within the optical spectrum, the redder wavelengths around $0.9-1$\um resemble an M4-type star. This is confirmed by comparing the spectral types derived from the equivalent widths (EW) of the CO 2.29\,\um absorption bandhead and the depth of the TiO 0.8859\,\um band. The EW of the CO bandhead is $\approx25$\,\angstrom, corresponding to an M7 spectral type \citep{Davies2007} in the $2-2.5$\,\um range, while the depth of the TiO band is $\approx0.25$, corresponding to an M3 spectral type \citep{Dorda2016} in the 0.8-0.9\,\um range. These comparisons suggest a transition from M1 to M7 spectral type from optical to NIR wavelengths. This mixed-temperature behavior persists in the later spectra from December 2024 and January 2025. 

Such mixed-temperature spectra have previously been observed in outbursting young stellar objects (YSO), where optical emission originates from the star and NIR emission from the accretion disk \citep{Rigliaco2012}. However, the high luminosity of WNTR23bzdiq rules out a YSO origin. Similar spectral behavior has been seen in the stellar merger eruption V838Mon \citep{Lynch04,Lynch07,Loebman15} and recently in the planet-star merger ZTF-SLRN-2020 \citep{De2023Nature}. Both sources underwent a transition from initial hot spectra to cool, M-type, mixed-temperature spectra at late times. In these events, the mixed-temperature spectrum is interpreted as arising from a multi-layered photosphere formed by the material ejected during the merger. For V\,838Mon, \citet{Lynch04, Lynch07} model the visible to mid-IR spectra using a configuration of a stellar blackbody viewed through dense outer shells. This configuration creates a complex, non-grey photosphere where longer wavelengths probe the outer regions of the envelope while shorter wavelengths probe the inner regions, producing the observed mixed-temperature spectrum. 

The mixed-temperature spectra of WNTR23bzdiq point towards a similar configuration of a dense envelope of ejecta around the star. Unlike V838\,Mon, where a clear transition from a hot progenitor to a cool M-type spectrum was observed, WNTR23bzdiq began as an M-type star. Nonetheless, the near-IR spectrum at the end of the first outburst indicates a transition to a slightly cooler spectral type than the progenitor. This is followed by a shift toward somewhat warmer M7-type spectra during the rebrightening phase. These spectral changes, though subtler than in V838\,Mon, reinforce the parallel with merger-driven eruptions. We further explore these similarities and the broader context of merger transients in Section \ref{sec:disc_ogle}. 

\section{Discussion}
\label{sec:discussion}
\subsection{The nature of WNTR23bzdiq}
We now discuss possible origins for the slow eruption of WNTR23bzdiq. AGB stars are known to exhibit brightness variations due to pulsations, being in symbiotic binaries, eclipses by companions, or due to thermal pulses. The variability in WNTR23bzdiq is also reminiscent of the class of ``slow variables" showing small luminosity changes over a span of several years (see \citealt{Sydney2025} for a recent summary). Known classes of slow variables include some YSOs, \citealp{Teixeira2018}, magnetically active main sequence (MS) and K-giant stars \citep{Baliunas1990, Phillips1978}, and precursors to stellar mergers \citep{Tylenda2011, Blagorodnova2020, Addison2022, Tranin2024}. However, this phase space of stellar variability has not been characterized extensively, and several examples exist that have not been fully explained. We examine each of these scenarios below.

First, WNTR23bzdiq is unlike the typical pulsation-driven variability seen in AGB stars \citep{Whitelock2003}. These stars exhibit periodic or semi-regular brightness variations on timescales ranging from a few hundred days to a few thousand days \citep{Whitelock2003, Riebel2015, Karambelkar2020}, unlike the eruptive nature of the WNTR23bzdiq variations. 

Second, cool giants in symbiotic binaries can exhibit erratic brightness variations driven by accretion events or dust formation episodes, often appearing as long-duration outbursts or dimmings \citep{De2022}. However, symbiotic stars typically exhibit strong emission lines of hydrogen and helium from ionized gas, which are absent in all spectra of WNTR23bzdiq. This makes a symbiotic binary origin unlikely.

Third, eclipses by stellar companions or debris disks produce long-duration dimming events in cool giants \citep{Rowan2021, Tzanidakis2023, Torres2022}. One possibility is that WNTR23bzdiq is brightening because it recently emerged from such a dimming phase. However, this would require the star to have been at its minimum state since at least 2009 (as no dimming events are seen in the PTF and NEOWISE lightcurves). Such a duration would exceed the previous record for dimming giants by over three years \citep{Tzanidakis2023}. The CFHT data further suggest either that the dimming lasted for at least 14 years or that the star underwent a brightening and fading episode between 2005 and 2008. The latter seems unlikely given the long brightening timescale observed post-2020 data, while the former would be unprecedented.

Fourth, while several known classes of slow variables such as YSOs, active MS or K-type giants are ruled out by the AGB progenitor of WNTR23bzdiq, some giant and AGB stars exhibit slow variations whose mechanisms are not fully understood. Recently, \citet{Sydney2025} cataloged slow variables from the ASAS-SN survey, noting that while most of the AGB stars exhibit large amplitude, semi-regular periodic variations, some giant stars exhibit steady brightening or dimming trends over several years. WNTR23bzdiq is more similar to this latter group and may represent a more evolved member. However, it does not closely resemble any of the individual examples of slow giants presented in their paper.\footnote{The closest resemblance is to the red giant NSV\,1610 (a.k.a. 2MASSJ04282152-0148343, their Fig. 11) which was at roughly constant brightness for six years (2014--2020) following which it underwent an 1 mag outburst for three years. However, the full lightcurve of this source from \href{http://asas-sn.ifa.hawaii.edu/skypatrol/objects/403727100335}{ASAS-SN Skypatrolv2} shows that before 2014, the star had a higher brightness, underwent a decline of 1.5 mag in 2012 and stayed at this low luminosity for the next 8 years. The 2020 outburst was a rebrightening to its initial brightness, unlike WNTR23bzdiq.}.

Fifth, we examined ZTF light curves for 1,700 red giants and supergiants in the outskirts of M31 compiled by \citet{Neugent2020}, and found no source with variability resembling WNTR23bzdiq. All variables in this sample show changes on timescales of a few hundred days, typical of standard RSG or AGB variability. We also searched the database of mid-IR transients in M31 from the NEOWISE survey \citep{De2024_m31} and found no comparable events. Taken together, these comparisons suggest that WNTR23bzdiq appears to be unlike known classes of slow variable stars.

Sixth, thermal pulses and hot-bottom burning can produce surface luminosity variations of up to $\sim$0.5\,dex in AGB stars \citep{Marigo2007, Marigo08, Marigo2013}. With $\approx10^{3}$ AGB stars in M31 brighter than $m_{H}<16.5$\,mag (similar to WNTR23bzdiq, \citealt{Gavetti2025}) and an interpulse duration of $\approx10^{4}$ years \citep{Marigo2007}, such events are expected roughly once per decade. It is therefore plausible that the observed brightening of WNTR23bzdiq is linked to a thermal pulse. However, thermal pulse-driven luminosity changes are predicted to occur over a timescale of $\sim10^{4}$ years, much slower than WNTR23bzdiq, whose luminosity increased by 0.5\,dex over a few years. Additionally, the \texttt{MIST} models suggest that the progenitor was on the early-AGB, before the onset of thermal pulses, making this explanation less likely. If this event is indeed a thermal pulse, WNTR23bzdiq would offer a rare opportunity to study this process. To our knowledge, the only other candidate is the donor star in the symbiotic binary IGR J17329-2731, which has brightened by 4 magnitudes over the last four decades, possibly due to a thermal pulse \citep{Bozzo2018}.

Finally, a compelling explanation links WNTR23bzdiq to eruptions associated with stellar mergers triggered by CEE. Binary systems undergoing CEE often exhibit slow, years-long brightenings from mass loss during the dynamical inspiral \citep{Tylenda11, Blagorodnova2020, Addison2022, Tranin2024}. When the primary is a giant or AGB star, the full merger is accompanied by a gradual increase in luminosity over several years \citep{Pejcha16b,MacLeod2022}, similar to that seen in WNTR23bzdiq. The spectroscopic similarities with the Galactic stellar merger V838\,Mon, discussed in Section \ref{sec:spec}, further support a merger association. In the following Section, we explore the resemblance between WNTR23bzdiq and the Galactic giant star merger OGLE-2002-BLG-360, and examine whether WNTR23bzdiq may mark the onset of CEE in an AGB star.

\subsection{WNTR23bzdiq: a slower analog of OGLE-2002-BLG-360?}
\label{sec:disc_ogle}
\begin{figure}
    \centering
    \includegraphics[width=0.44\textwidth]{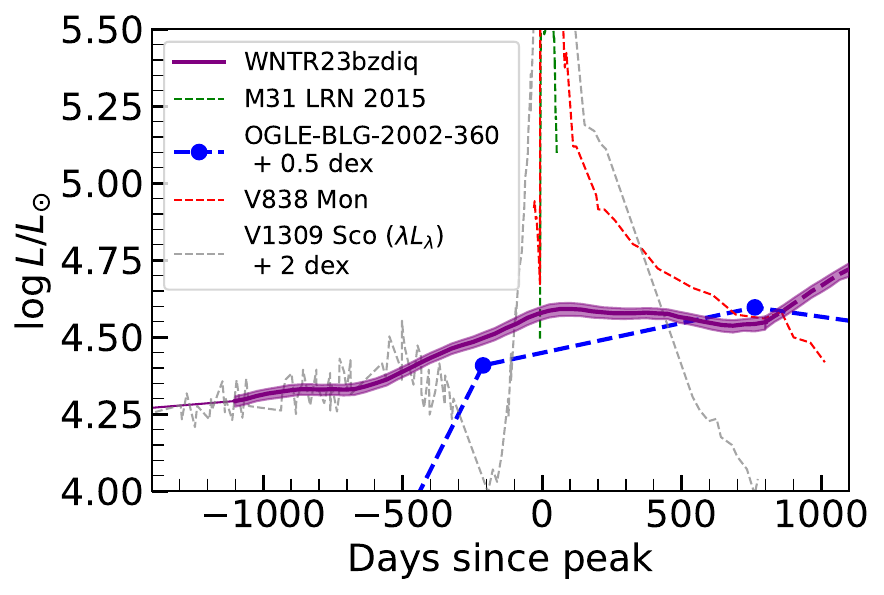}\\
    \includegraphics[width=0.44\textwidth]{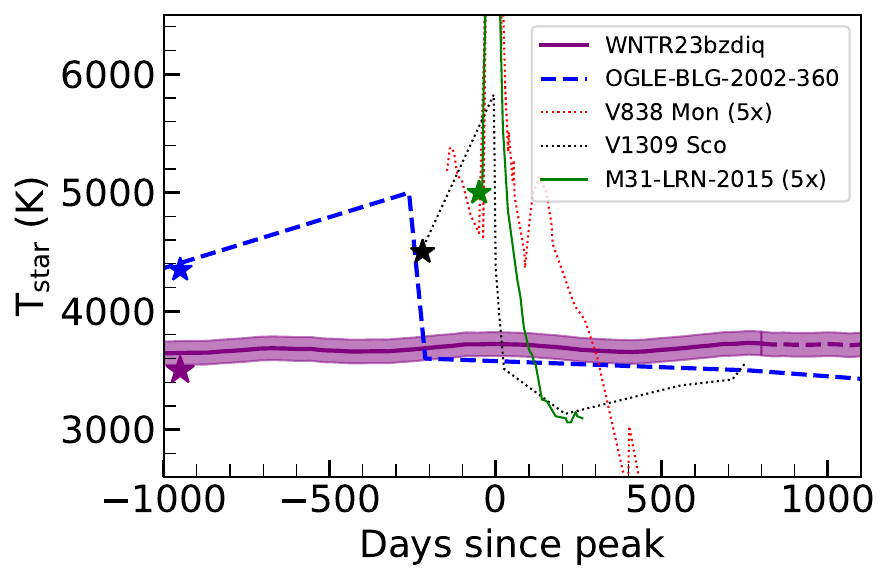}\\
    \includegraphics[width=0.44\textwidth]{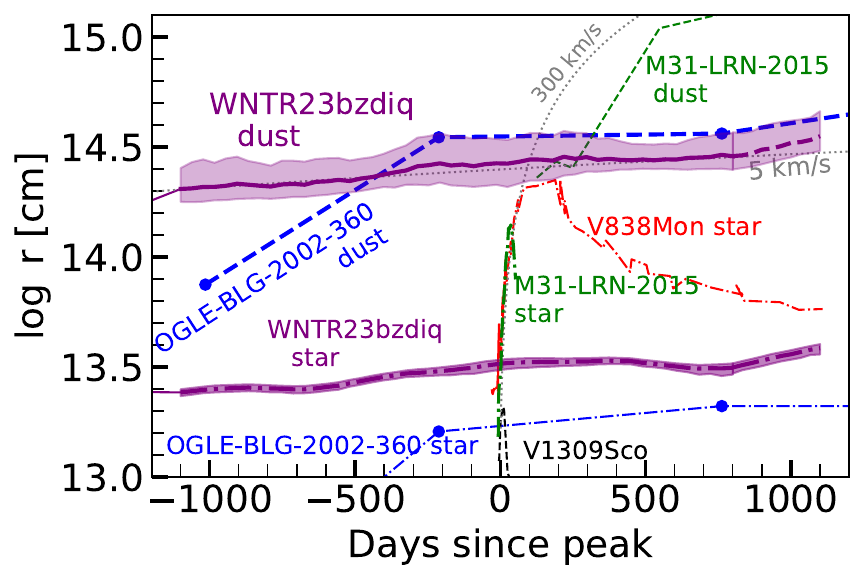}     \caption{Comparison of properties of WNTR23bzdiq (purple, reference MJD 59500) to the giant star merger OGLE-BLG (blue, MJD 53125) and classical LRNe V838\,Mon (red, MJD 52315), V1309\,Sco (black, MJD 54725), and M31-LRN-2015 (green, MJD 57044). The phases are plotted relative to MJD=59500 for WNTR23bzdiq, \emph{Top} The classical LRNe are relatively short lived-eruptions with a sharp rise to maximum, while OGLE-BLG shows a slower rise and longer duration than the classical LRNe. WNTR23bzdiq has an even slower rise and longer duration, consistent with its progenitor being more evolved and extended than that of OGLE-BLG. \emph{Middle:} During their eruptions, all stellar mergers become hotter than their progenitors (indicated by stars) at initial times and then cool rapidly to resemble cold M-type stars. The progenitor of WNTR23bzdiq is a M-type star to begin with, and so it is not expected to show these sharp transitions. The long term temperature is similar to that reached by OGLE-BLG. \emph{Bottom:} The LRNe show rapid photospheric and dust-shell expansions with velocities of $\approx300$\,km\,s$^{-1}$ (black dotted line). In contrast, WNTR23bzdiq and OGLE-BLG show a much slower radius evolution with dust-radius expansion velocities of $\approx5$\,km\,s$^{-1}$.} 
    \label{fig:property_comparisons}
\end{figure}

The Galactic transient OGLE-2002-BLG-360 (OGLE-BLG, \citealt{Tylenda13}) is considered to be the prototypical (and the only unambiguous) example of a giant star merger. Its eruption began in 2002 with a four magnitude rise over five hundred days, followed by three lightcurve peaks and a transition to an M-type supergiant, consistent with the hallmarks of stellar mergers such as V838\,Mon and V1309\,Sco. However, OGLE-BLG's eruption spanned $\approx$2000 days -- over an order of magnitude longer than the other mergers -- likely due to its extended giant primary, unlike the MS progenitors of V838\,Mon and V1309\,Sco \citep{MacLeod2022}. The orbital period at the surface of the progenitor of OGLE-BLG (M$\approx1-3$\,M$_{\odot}$, R$\approx 30$\,R$_{\odot}$, T$_{\rm{eff}}\approx4350$\,K) is $\approx 19$ days -- substantially longer than the other mergers (e.g. 1.4 days for V1309\,Sco, \citealt{Tylenda11}). The slow bumpy eruption of WNTR23bzdiq is reminiscent of OGLE-BLG, though with even longer timescales. This is expected in the merger scenario, as the AGB progenitor of WNTR23bzdiq is more extended than that of OGLE-BLG (orbital period at surface $\approx270$\,days), and would produce an even slower merger than OGLE-BLG. Motivated by this, we examine WNTR23bzdiq as a possible analog of OGLE-BLG-2002-360.

Figure \ref{fig:property_comparisons} compares the luminosities, temperatures and radii of WNTR23bzdiq with OGLE-BLG \citep{Tylenda13} and the more classical stellar mergers (or LRNe) V838Mon \citep{Munari2002AA, Tylenda05a}, V1309Sco \citep{Tylenda11}, and M31-LRN-2015 \citep{MacLeod17, Blagorodnova2020}. The luminosity evolution (top panel) of WNTR23bzdiq qualitatively resembles that of OGLE-BLG, but has a lower amplitude (a factor of 2 increase versus 40) and is stretched out over a longer duration. Both WNTR23bzdiq and OGLE-BLG evolve over much longer timescales than the main eruptions of the three LRNe. However, WNTR23bzdiq resembles the slow \emph{I}-band rise observed in V1309\,Sco for several years before its main eruption, which has been attributed to pre-merger mass loss from the binary \citep{Pejcha14}. 

The middle panel shows the evolution of the stellar effective temperatures. All comparison LRNe had progenitors warmer than 4000\,K (marked by stars), with temperatures that typically rose at the onset of the eruption before rapidly dropping to $\leq3500$\,K as they evolved into M-type stars. In contrast, WNTR23bzdiq was an M-type star to start with, and continued to stay at $\approx 3500-3600$\,K without any significant changes throughout its eruption. OGLE-BLG (blue dashed line) started out warmer (at 4300\,K) but reached a similar temperature of $\approx3500$\,K at late times. 

The bottom panel compares the evolution of the photospheric and dust shell radii. Both the stellar and dust shell radii of WNTR23bzdiq closely track those of OGLE-BLG during its eruption, with similarly slow expansion velocities of $\approx 5$\,km\,s$^{-1}$. In contrast, classical LRNe exhibit much faster expansion, around $\approx300$\,km\,s$^{-1}$. The slow expansion velocities for OGLE-BLG and WNTR23bzdiq can be explained by a slow wind from the stellar surface over the course of its eruption, maintaining the dust shell at roughly the same radius \citep{Steinmetz2025}, unlike LRNe that are marked by a rapid mass-ejection episode at the time of merger which expands with high velocities. This is in line with the expectation that CE stripping in extended giants occurs as a steady wind over several years, rather than a sudden mass ejection seen in less extended stars \citep{MacLeod2022}.

Thus, WNTR23bzdiq shares several similarities with OGLE-BLG, appearing as a scaled-down, stretched-out analog, consistent with its more extended, cooler M-type AGB progenitor compared to OGLE-BLG. To assess the plausibility of a CE origin for WNTR23bzdiq, we compare the energetics of the eruption and with the orbital energy losses. 

The total radiated energy of WNTR23bzdiq is $\approx 10^{46}$\,erg. The magnitude of the orbital energy of a $\sim$1\,M$_{\odot}$ companion at the surface of the progenitor star is $\sim10^{47}$\,erg, suggesting that $\sim10$\% of the orbital energy has been radiated if the companion has not inspiraled deeper within the envelope. If the stellar orbit has shrunk by a factor of ten, only $\sim$1\% of the lost orbital energy needs to be radiated. In this case, the lost orbital energy would approach the envelope's binding energy ($\sim10^{48}$\,erg), making envelope ejection possible. In the extreme case of a full merger, the radiated energy corresponds to $\sim$0.1\% of the orbital energy lost. 

We estimate the expected luminosity from a CE inspiral using the formalism described in \citep{MacLeod2018}. If a  $\sim1$\,M$_{\odot}$ companion spirals through the progenitor's outer envelope, the rate of orbital energy loss (L$_{\rm{decay}}$) is $\sim 10^{39}$\,erg\,s$^{-1}$. The peak radiated luminosity in WNTR23bzdiq is $\sim10^{38}$\,erg\,s$^{-1}$ -- about 10\% of L$_{\rm{decay}}$. Since L$_{\rm{decay}} \propto R^{-5/2}$, the orbital energy is lost more rapidly if the companion is embedded deeper within the envelope. In all cases, these estimates suggest that only a fraction of the orbital energy is required to power the observed luminosity and radiated energy of WNTR23bzdiq. Future observations of this ongoing eruption will determine whether the energetics remain consistent with the CE explanation.

In summary, WNTR23bzdiq appears to be a slower analog of OGLE-BLG, consistent with a cooler, more extended AGB progenitor. The multi-peaked, slow eruption is ongoing and provides a valuable opportunity for continued multiwavelength follow-up. 

\begin{figure*}[hbt]
    \centering
    \includegraphics[width=0.7\textwidth]{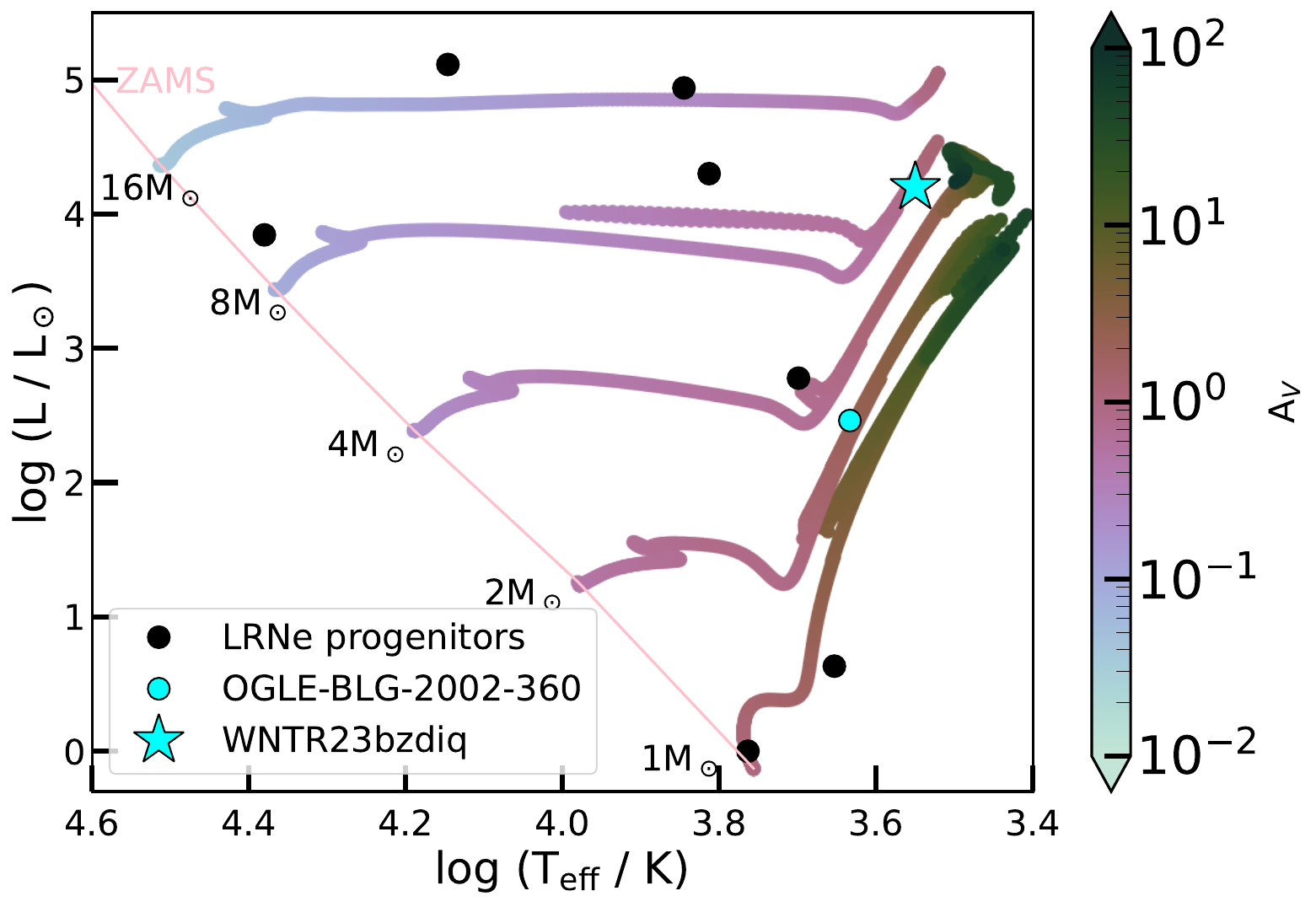}
    \caption{HR-diagram with the progenitors of classical LRNe (black dots), OGLE-BLG (cyan dot) and WNTR23bzdiq (cyan star) with MIST stellar evolutionary tracks. The stellar tracks are color-coded with the expected dust obscurations (A$_{V}$) from the wind-model of \citet{MacLeod2022}. The progenitor of WNTR23bzdiq is colder than that of any other merger transient, and is the only progenitor on the AGB -- potentially providing a unique opportunity to study the outcomes of common-envelope evolution in an AGB star.}
    \label{fig:lrn_prog_hrd}
\end{figure*}

\subsection{WNTR23bzdiq in the context of common-envelope transients}
Figure \ref{fig:lrn_prog_hrd} shows the HR diagram locations of the progenitor primary stars  of WNTR23bzdiq and other LRNe. The AGB progenitor of WNTR23bzdiq (cyan star) is cooler and more evolved than the MS or Hertzsprung-Gap progenitors of all LRNe. As noted before, giant star primaries are rare in this sample, with OGLE-BLG being the only other example (cyan dot). This scarcity has been attributed to large dust obscuration due to pre-merger mass loss in such systems \citep{MacLeod2022}. Overlaid stellar evolutionary tracks in Figure \ref{fig:lrn_prog_hrd} (colored by dust obscuration from the wind model of \citealt{MacLeod2022}) show that for a 1\,M$_{\odot}$ companion, the WNTR23bzdiq progenitor would have $A_{V}\approx 0.5-0.7$, slightly lower than the value of $A_{V}\approx 2 - 3$ from our \texttt{DUSTY} modeling. This discrepancy could reflect a more massive companion ($\sim3$\,M$_{\odot}$), or the limitations of the \texttt{DUSTY} modeling discussed in Section \ref{sec:progenitor}. In any case, WNTR23bzdiq sits at an intermediate level of dust obscuration between MS/HG LRN progenitors (A$_{V}\leq0.1$) and low-mass giants and AGB stars (A$_{V}>10$\,mag). While optical time-domain surveys will continue to discover more LRNe, upcoming IR time-domain searches will be crucial to probe the dustier transients. 

In the CE scenario, the extended progenitor of WNTR23bzdiq makes it a candidate for a successful envelope ejection rather than a stellar merger. LRN progenitors typically have radiative envelopes with high binding energies and are more likely to merge with their companions after CEE, while a successful CE ejection requires a cool, extended envelope with lower binding energies \citep{Klencki2021, MacLeod2022}. Our estimates for WNTR23bzdiq show that the orbital energy lost exceeds the binding energy of the envelope if the inspiral shrinks the orbit by a factor of ten. Future observations to measure the mass of the ejected material will help clarify the outcome of this event. 

The properties of WNTR23bzdiq are also consistent with theoretical simulations of CEE. The multiple bumps in its lightcurve could arise from mild shocks that form when envelope layers ejected at later times collide with the previously ejected layers \citep{Ricker2012, Iaconi2018, Iaconi2019}. In this picture, assuming that the ejecta travel at the escape velocity of the primary star ($\approx88$\,km\,s$^{-1}$), the ejected mass needs to be $\sim$0.1\,M$_{\odot}$ to have sufficiently large kinetic energy to explain the energy radiated in the second peak. Alternatively, the rebrightening could be powered by instabilities during the slow self-regulated phase that produce large amplitude pulsations in the envelope \citep{Clayton2017}; or by accretion episodes that could produce fast-moving ($>100$\,km\,s$^{-1}$) jets, possibly resulting in the formation of bipolar planetary nebulae \citep{Soker2012, Soker2020}. Our observations only trace the slowly expanding photosphere and not the actual ejecta, and we cannot constrain the speeds of the fastest-moving ejecta in WNTR23bzdiq due to a lack of emission or P-Cygni lines. Continued spectroscopic observations to search for emission lines from the fast-moving ejecta will shed light on the role of jets in this event. The current dust mass ($\sim10^{-7}$\,M$_{\odot}$) is broadly consistent with early-phase predictions from CEE simulations \citep{Gonzalez-Bolivar2024, Bermudez2024, Iaconi2020, Lu2013}, which predict substantial dust formation (10$^{-3}-10^{-2}$\,M$_{\odot}$) only $\sim10$ years after the onset of CEE. Future \emph{JWST} observations to trace the evolution of the dust content will be key to confirming the CE origin for WNTR23bzdiq.

\section{Summary and way forward}
\label{sec:summary}
In this paper, we presented WNTR23bzdiq / WTP19aalzlk --- a slow eruption of an early-AGB star in the M\,31 galaxy identified by the WINTER near-infrared and the NEOWISE mid-infrared surveyors, that exhibits the following properties --- 

\begin{itemize}[leftmargin=*]
    \item The source showed a slow, bumpy increase in its optical flux beginning in 2018. From 2018 to 2021, it steadily brightened by 0.5\,mag in the \emph{r}-band. In 2021, it underwent a sharp outburst with an amplitude of 1\,mag lasting for a total of a thousand days. In 2024, the source rebrightened by 1\,mag and has continued to rise since. At the time of writing, the source has brightened by 2 magnitudes since its quiescent state in the last ten years. 
    \item The optical outburst was accompanied by a slow mid-IR brightening (1\,mag over ten years) seen in the NEOWISE data. The mid-IR flux increases smoothly in this duration and continued to rise even when the optical flux faded in 2022. 
    \item No substantial photometric variations were seen in the optical bands from 2003 to 2015. Modeling the quiescent data suggests that the progenitor was an early-AGB star with a L\,$\approx$ 16000\,L$_{\odot}$ and T$_{\rm{eff}}\approx3500$\,K.
    \item The bolometric luminosity rose to $\approx40000$\,L$_{\odot}$ during the 2021 eruption, plateaued for the next two years, and has risen to $\approx50000$\,L$_{\odot}$ since then. The temperature stayed mostly constant at $\approx3600$\,K throughout the eruption. The stellar and dust shell radii suggest slow photospheric expansion velocities of $\approx 5$\,km\,s$^{-1}$, indicating a prolonged mass-loss event. The dust mass produced thus far is $\sim4\times10^{-7}$\,M$_{\odot}$.
    \item Optical and NIR spectra taken during the outburst show strong molecular absorption features of TiO and resemble late M-type stars. The spectra exhibit mixed-temperature behavior where the spectral type transitions from M1 in the optical to M7/8 in the NIR, similar to that seen in the Galactic stellar merger V838\,Mon. 
\end{itemize}

WNTR23bzdiq is unlike known classes of slow giant variables such as pulsating AGB stars, symbiotic binaries, and giant star eclipses. While its variability could be triggered by a thermal pulse in an isolated AGB star, the observed timescales are significantly faster than those predicted by theoretical models. The spectroscopic and photometric properties also resemble those of stellar merger transients -- particularly the giant star merger OGLE-2002-BLG-360 -- but on much longer timescales, consistent with those expected from its colder and more extended AGB progenitor. As such, the slow eruption of WNTR23bzdiq possibly marks the onset of CEE in a binary with an AGB primary, providing a rare opportunity to study CEE in such systems. Continued monitoring --- particularly mid-IR observations with \emph{JWST} to trace the evolution of dust masses and multiband observations to measure the energetics of the eruption --- will help confirm the association with CEE. 

WNTR23bzdiq is possibly a member of the emerging class of slow, infrared-bright transients from CEE in cold giant primaries, whose extended envelopes have low binding energies, making them more likely to eject their envelopes as a consequence of CEE. New and upcoming infrared transient searches such as WINTER, PRime Focus Infrared Microlensing Experiment (PRIME, \citealt{Konndo2023}), Dynamic Red All-sky Monitoring Survey (DREAMS \citealt{Tranin2024}), and Cryoscope \citep{Kasliwal2025_cryoscope} will be instrumental in mapping the largely uncharted landscape of such transients. These surveys will set the stage for more sensitive infrared searches with the \emph{Nancy Grace Roman Space Telescope}. 

\section*{Acknowledgements}
VK acknowledges Yashvi Sharma for assistance with the machine learning interpolator, and Jacob Jencson for helping with the GRAMS modeling. WINTER’s construction is made possible by the National Science Foundation under MRI grant number AST-1828470 with early operations supported by AST-2206730. Significant support for WINTER also comes from the California Institute of Technology, the Caltech Optical Observatories, the Bruno Rossi Fund of the MIT Kavli Institute for Astrophysics and Space Research, the David and Lucille Packard Foundation, and the MIT Department of Physics and School of Science. This work was supported by a NASA Keck PI Data Award, administered by the NASA Exoplanet Science Institute. We acknowledge the support of the National Aeronautics and Space Administration through ADAP grant number 80NSSC24K0663. This research has made use of the Keck Observatory Archive (KOA), which is operated by the W. M. Keck Observatory and the NASA Exoplanet Science Institute (NExScI), under contract with the National Aeronautics and Space Administration. N. B. acknowledges to be funded by the European Union (ERC, CET-3PO, 101042610). Views and opinions expressed are however those of the author(s) only and do not necessarily reflect those of the European Union or the European Research Council Executive Agency. Neither the European Union nor the granting authority can be held responsible for them. Some of the data presented herein were obtained at Keck Observatory, which is a private 501(c)3 non-profit organization operated as a scientific partnership among the California Institute of Technology, the University of California, and the National Aeronautics and Space Administration. The Observatory was made possible by the generous financial support of the W. M. Keck Foundation. The authors wish to recognize and acknowledge the very significant cultural role and reverence that the summit of Maunakea has always had within the Native Hawaiian community. We are most fortunate to have the opportunity to conduct observations from this mountain.

\bibliography{myreferences}

\begin{thebibliography}{}
\expandafter\ifx\csname natexlab\endcsname\relax\def\natexlab#1{#1}\fi
\providecommand{\url}[1]{\href{#1}{#1}}
\providecommand{\dodoi}[1]{doi:~\href{http://doi.org/#1}{\nolinkurl{#1}}}
\providecommand{\doeprint}[1]{\href{http://ascl.net/#1}{\nolinkurl{http://ascl.net/#1}}}
\providecommand{\doarXiv}[1]{\href{https://arxiv.org/abs/#1}{\nolinkurl{https://arxiv.org/abs/#1}}}

\bibitem[{{Addison} {et~al.}(2022){Addison}, {Blagorodnova}, {Groot}, {Erasmus}, {Jones}, \& {Mogawana}}]{Addison2022}
{Addison}, H., {Blagorodnova}, N., {Groot}, P.~J., {et~al.} 2022, \mnras, 517, 1884, \dodoi{10.1093/mnras/stac2685}

\bibitem[{Ansel {et~al.}(2024)Ansel, Yang, He, Gimelshein, Jain, Voznesensky, Bao, Bell, Berard, Burovski, Chauhan, Chourdia, Constable, Desmaison, DeVito, Ellison, Feng, Gong, Gschwind, Hirsh, Huang, Kalambarkar, Kirsch, Lazos, Lezcano, Liang, Liang, Lu, Luk, Maher, Pan, Puhrsch, Reso, Saroufim, Siraichi, Suk, Suo, Tillet, Wang, Wang, Wen, Zhang, Zhao, Zhou, Zou, Mathews, Chanan, Wu, \& Chintala}]{pytorch}
Ansel, J., Yang, E., He, H., {et~al.} 2024, in 29th ACM International Conference on Architectural Support for Programming Languages and Operating Systems, Volume 2 (ASPLOS '24) (ACM), \dodoi{10.1145/3620665.3640366}

\bibitem[{{Baliunas} \& {Jastrow}(1990)}]{Baliunas1990}
{Baliunas}, S., \& {Jastrow}, R. 1990, \nat, 348, 520, \dodoi{10.1038/348520a0}

\bibitem[{{Bellm} {et~al.}(2019){Bellm}, {Kulkarni}, {Graham}, {Dekany}, {Smith}, {Riddle}, {Masci}, {Helou}, {Prince}, {Adams}, {Barbarino}, {Barlow}, {Bauer}, {Beck}, {Belicki}, {Biswas}, {Blagorodnova}, {Bodewits}, {Bolin}, {Brinnel}, {Brooke}, {Bue}, {Bulla}, {Burruss}, {Cenko}, {Chang}, {Connolly}, {Coughlin}, {Cromer}, {Cunningham}, {De}, {Delacroix}, {Desai}, {Duev}, {Eadie}, {Farnham}, {Feeney}, {Feindt}, {Flynn}, {Franckowiak}, {Frederick}, {Fremling}, {Gal-Yam}, {Gezari}, {Giomi}, {Goldstein}, {Golkhou}, {Goobar}, {Groom}, {Hacopians}, {Hale}, {Henning}, {Ho}, {Hover}, {Howell}, {Hung}, {Huppenkothen}, {Imel}, {Ip}, {Ivezi{\'c}}, {Jackson}, {Jones}, {Juric}, {Kasliwal}, {Kaspi}, {Kaye}, {Kelley}, {Kowalski}, {Kramer}, {Kupfer}, {Landry}, {Laher}, {Lee}, {Lin}, {Lin}, {Lunnan}, {Giomi}, {Mahabal}, {Mao}, {Miller}, {Monkewitz}, {Murphy}, {Ngeow}, {Nordin}, {Nugent}, {Ofek}, {Patterson}, {Penprase}, {Porter}, {Rauch}, {Rebbapragada}, {Reiley}, {Rigault}, {Rodriguez}, {van Roestel}, {Rusholme}, {van
  Santen}, {Schulze}, {Shupe}, {Singer}, {Soumagnac}, {Stein}, {Surace}, {Sollerman}, {Szkody}, {Taddia}, {Terek}, {Van Sistine}, {van Velzen}, {Vestrand}, {Walters}, {Ward}, {Ye}, {Yu}, {Yan}, \& {Zolkower}}]{Bellm2019}
{Bellm}, E.~C., {Kulkarni}, S.~R., {Graham}, M.~J., {et~al.} 2019, \pasp, 131, 018002, \dodoi{10.1088/1538-3873/aaecbe}

\bibitem[{{Berm{\'u}dez-Bustamante} {et~al.}(2024){Berm{\'u}dez-Bustamante}, {De Marco}, {Siess}, {Price}, {Gonz{\'a}lez-Bol{\'\i}var}, {Lau}, {Mu}, {Hirai}, {Danilovich}, \& {Kasliwal}}]{Bermudez2024}
{Berm{\'u}dez-Bustamante}, L.~C., {De Marco}, O., {Siess}, L., {et~al.} 2024, \mnras, 533, 464, \dodoi{10.1093/mnras/stae1841}

\bibitem[{{Blagorodnova} {et~al.}(2017){Blagorodnova}, {Kotak}, {Polshaw}, {Kasliwal}, {Cao}, {Cody}, {Doran}, {Elias-Rosa}, {Fraser}, {Fremling}, {Gonzalez-Fernandez}, {Harmanen}, {Jencson}, {Kankare}, {Kudritzki}, {Kulkarni}, {Magnier}, {Manulis}, {Masci}, {Mattila}, {Nugent}, {Ochner}, {Pastorello}, {Reynolds}, {Smith}, {Sollerman}, {Taddia}, {Terreran}, {Tomasella}, {Turatto}, {Vreeswijk}, {Wozniak}, \& {Zaggia}}]{Blagorodnova17}
{Blagorodnova}, N., {Kotak}, R., {Polshaw}, J., {et~al.} 2017, \apj, 834, 107, \dodoi{10.3847/1538-4357/834/2/107}

\bibitem[{{Blagorodnova} {et~al.}(2020){Blagorodnova}, {Karambelkar}, {Adams}, {Kasliwal}, {Kochanek}, {Dong}, {Campbell}, {Hodgkin}, {Jencson}, {Johansson}, {Koz{\l}owski}, {Laher}, {Masci}, {Nugent}, \& {Rebbapragada}}]{Blagorodnova2020}
{Blagorodnova}, N., {Karambelkar}, V., {Adams}, S.~M., {et~al.} 2020, \mnras, 496, 5503, \dodoi{10.1093/mnras/staa1872}

\bibitem[{{Blagorodnova} {et~al.}(2021){Blagorodnova}, {Klencki}, {Pejcha}, {Vreeswijk}, {Bond}, {Burdge}, {De}, {Fremling}, {Gehrz}, {Jencson}, {Kasliwal}, {Kupfer}, {Lau}, {Masci}, \& {Rich}}]{Blagorodnova2021}
{Blagorodnova}, N., {Klencki}, J., {Pejcha}, O., {et~al.} 2021, \aap, 653, A134, \dodoi{10.1051/0004-6361/202140525}

\bibitem[{{Bozzo} {et~al.}(2018){Bozzo}, {Bahramian}, {Ferrigno}, {Sanna}, {Strader}, {Lewis}, {Russell}, {di Salvo}, {Burderi}, {Riggio}, {Papitto}, {Gandhi}, \& {Romano}}]{Bozzo2018}
{Bozzo}, E., {Bahramian}, A., {Ferrigno}, C., {et~al.} 2018, \aap, 613, A22, \dodoi{10.1051/0004-6361/201832588}

\bibitem[{{Cai} {et~al.}(2022){Cai}, {Pastorello}, {Fraser}, {Wang}, {Filippenko}, {Reguitti}, {Patra}, {Goranskij}, {Barsukova}, {Brink}, {Elias-Rosa}, {Stevance}, {Zheng}, {Yang}, {Atapin}, {Benetti}, {de Boer}, {Bose}, {Burke}, {Byrne}, {Cappellaro}, {Chambers}, {Chen}, {Emami}, {Gao}, {Hiramatsu}, {Howell}, {Huber}, {Kankare}, {Kelly}, {Kotak}, {Kravtsov}, {Lander}, {Li}, {Lin}, {Lundqvist}, {Magnier}, {Malygin}, {Maslennikova}, {Matilainen}, {Mazzali}, {McCully}, {Mo}, {Moran}, {Newsome}, {Oparin}, {Padilla Gonzalez}, {Reynolds}, {Shatsky}, {Smartt}, {Smith}, {Stritzinger}, {Tatarnikov}, {Terreran}, {Uklein}, {Valerin}, {Vallely}, {Vozyakova}, {Wainscoat}, {Yan}, {Zhang}, {Zhang}, {Zheltoukhov}, {Dastidar}, {Fulton}, {Galbany}, {Gangopadhyay}, {Ge}, {Guti{\'e}rrez}, {Lin}, {Misra}, {Ou}, {Salmaso}, {Tartaglia}, {Xiao}, \& {Zhang}}]{Cai2022}
{Cai}, Y.~Z., {Pastorello}, A., {Fraser}, M., {et~al.} 2022, arXiv e-prints, arXiv:2207.00734.
\newblock \doarXiv{2207.00734}

\bibitem[{{Cao} {et~al.}(2016){Cao}, {Nugent}, \& {Kasliwal}}]{Cao16}
{Cao}, Y., {Nugent}, P.~E., \& {Kasliwal}, M.~M. 2016, \pasp, 128, 114502, \dodoi{10.1088/1538-3873/128/969/114502}

\bibitem[{{Chambers} {et~al.}(2016){Chambers}, {Magnier}, {Metcalfe}, {Flewelling}, {Huber}, {Waters}, {Denneau}, {Draper}, {Farrow}, {Finkbeiner}, {Holmberg}, {Koppenhoefer}, {Price}, {Saglia}, {Schlafly}, {Smartt}, {Sweeney}, {Wainscoat}, {Burgett}, {Grav}, {Heasley}, {Hodapp}, {Jedicke}, {Kaiser}, {Kudritzki}, {Luppino}, {Lupton}, {Monet}, {Morgan}, {Onaka}, {Stubbs}, {Tonry}, {Banados}, {Bell}, {Bender}, {Bernard}, {Botticella}, {Casertano}, {Chastel}, {Chen}, {Chen}, {Cole}, {Deacon}, {Frenk}, {Fitzsimmons}, {Gezari}, {Goessl}, {Goggia}, {Goldman}, {Grebel}, {Hambly}, {Hasinger}, {Heavens}, {Heckman}, {Henderson}, {Henning}, {Holman}, {Hopp}, {Ip}, {Isani}, {Keyes}, {Koekemoer}, {Kotak}, {Long}, {Lucey}, {Liu}, {Martin}, {McLean}, {Morganson}, {Murphy}, {Nieto-Santisteban}, {Norberg}, {Peacock}, {Pier}, {Postman}, {Primak}, {Rae}, {Rest}, {Riess}, {Riffeser}, {Rix}, {Roser}, {Schilbach}, {Schultz}, {Scolnic}, {Szalay}, {Seitz}, {Shiao}, {Small}, {Smith}, {Soderblom}, {Taylor}, {Thakar}, {Thiel},
  {Thilker}, {Urata}, {Valenti}, {Walter}, {Watters}, {Werner}, {White}, {Wood-Vasey}, \& {Wyse}}]{Chambers16}
{Chambers}, K.~C., {Magnier}, E.~A., {Metcalfe}, N., {et~al.} 2016, arXiv:1612.05560.
\newblock \doarXiv{1612.05560}

\bibitem[{{Choi} {et~al.}(2016){Choi}, {Dotter}, {Conroy}, {Cantiello}, {Paxton}, \& {Johnson}}]{Choi2016}
{Choi}, J., {Dotter}, A., {Conroy}, C., {et~al.} 2016, \apj, 823, 102, \dodoi{10.3847/0004-637X/823/2/102}

\bibitem[{{Clayton} {et~al.}(2017){Clayton}, {Podsiadlowski}, {Ivanova}, \& {Justham}}]{Clayton2017}
{Clayton}, M., {Podsiadlowski}, P., {Ivanova}, N., \& {Justham}, S. 2017, \mnras, 470, 1788, \dodoi{10.1093/mnras/stx1290}

\bibitem[{Coughlin {et~al.}(2023)Coughlin, Bloom, Nir, Antier, du~Laz, van~der Walt, Crellin-Quick, Culino, Duev, Goldstein, Healy, Karambelkar, Lilleboe, Shin, Singer, Ahumada, Anand, Bellm, Dekany, Graham, Kasliwal, Kostadinova, Kiendrebeogo, Kulkarni, Jenkins, LeBaron, Mahabal, Neill, Parazin, Peloton, Perley, Riddle, Rusholme, van Santen, Sollerman, Stein, Turpin, Wold, Amat, Bonnefon, Bonnefoy, Flament, Kerkow, Kishore, Jani, Mahanty, Liu, Llinares, Makarison, Olliéric, Perez, Pont, \& Sharma}]{Coughlin2023}
Coughlin, M.~W., Bloom, J.~S., Nir, G., {et~al.} 2023, The Astrophysical Journal Supplement Series, 267, 31, \dodoi{10.3847/1538-4365/acdee1}

\bibitem[{{Cushing} {et~al.}(2004){Cushing}, {Vacca}, \& {Rayner}}]{Cushing2004}
{Cushing}, M.~C., {Vacca}, W.~D., \& {Rayner}, J.~T. 2004, \pasp, 116, 362, \dodoi{10.1086/382907}

\bibitem[{{Cutri} {et~al.}(2003){Cutri}, {Skrutskie}, {van Dyk}, {Beichman}, {Carpenter}, {Chester}, {Cambresy}, {Evans}, {Fowler}, {Gizis}, {Howard}, {Huchra}, {Jarrett}, {Kopan}, {Kirkpatrick}, {Light}, {Marsh}, {McCallon}, {Schneider}, {Stiening}, {Sykes}, {Weinberg}, {Wheaton}, {Wheelock}, \& {Zacarias}}]{Cutri2003}
{Cutri}, R.~M., {Skrutskie}, M.~F., {van Dyk}, S., {et~al.} 2003, {VizieR Online Data Catalog: 2MASS All-Sky Catalog of Point Sources (Cutri+ 2003)}, VizieR On-line Data Catalog: II/246. Originally published in: University of Massachusetts and Infrared Processing and Analysis Center, (IPAC/California Institute of Technology) (2003)

\bibitem[{{Cutri} {et~al.}(2021){Cutri}, {Wright}, {Conrow}, {Fowler}, {Eisenhardt}, {Grillmair}, {Kirkpatrick}, {Masci}, {McCallon}, {Wheelock}, {Fajardo-Acosta}, {Yan}, {Benford}, {Harbut}, {Jarrett}, {Lake}, {Leisawitz}, {Ressler}, {Stanford}, {Tsai}, {Liu}, {Helou}, {Mainzer}, {Gettngs}, {Gonzalez}, {Hoffman}, {Marsh}, {Padgett}, {Skrutskie}, {Beck}, {Papin}, \& {Wittman}}]{Cutri2014}
{Cutri}, R.~M., {Wright}, E.~L., {Conrow}, T., {et~al.} 2021, {VizieR Online Data Catalog: AllWISE Data Release (Cutri+ 2013)}, VizieR On-line Data Catalog: II/328. Originally published in: IPAC/Caltech (2013)

\bibitem[{{Davies} {et~al.}(2007){Davies}, {Figer}, {Kudritzki}, {MacKenty}, {Najarro}, \& {Herrero}}]{Davies2007}
{Davies}, B., {Figer}, D.~F., {Kudritzki}, R.-P., {et~al.} 2007, \apj, 671, 781, \dodoi{10.1086/522224}

\bibitem[{{De} {et~al.}(2022){De}, {Mereminskiy}, {Soria}, {Conroy}, {Kara}, {Anand}, {Ashley}, {Boyer}, {Chakrabarty}, {Grefenstette}, {Hankins}, {Hillenbrand}, {Jencson}, {Karambelkar}, {Kasliwal}, {Lau}, {Lutovinov}, {Moore}, {Ng}, {Panagiotou}, {Pasham}, {Semena}, {Simcoe}, {Soon}, {Srinivasaragavan}, {Travouillon}, \& {Yao}}]{De2022}
{De}, K., {Mereminskiy}, I., {Soria}, R., {et~al.} 2022, \apj, 935, 36, \dodoi{10.3847/1538-4357/ac7c6e}

\bibitem[{{De} {et~al.}(2023){De}, {MacLeod}, {Karambelkar}, {Jencson}, {Chakrabarty}, {Conroy}, {Dekany}, {Eilers}, {Graham}, {Hillenbrand}, {Kara}, {Kasliwal}, {Kulkarni}, {Lau}, {Loeb}, {Masci}, {Medford}, {Meisner}, {Patel}, {Quiroga-Nu{\~n}ez}, {Riddle}, {Rusholme}, {Simcoe}, {Sjouwerman}, {Teague}, \& {Vanderburg}}]{De2023Nature}
{De}, K., {MacLeod}, M., {Karambelkar}, V., {et~al.} 2023, \nat, 617, 55, \dodoi{10.1038/s41586-023-05842-x}

\bibitem[{{De} {et~al.}(2024){De}, {MacLeod}, {Jencson}, {Lovegrove}, {Antoni}, {Kara}, {Kasliwal}, {Lau}, {Loeb}, {Masterson}, {Meisner}, {Panagiotou}, {Quataert}, \& {Simcoe}}]{De2024_m31}
{De}, K., {MacLeod}, M., {Jencson}, J.~E., {et~al.} 2024, arXiv e-prints, arXiv:2410.14778, \dodoi{10.48550/arXiv.2410.14778}

\bibitem[{{Dominik} {et~al.}(2012){Dominik}, {Belczynski}, {Fryer}, {Holz}, {Berti}, {Bulik}, {Mandel}, \& {O'Shaughnessy}}]{Dominik12}
{Dominik}, M., {Belczynski}, K., {Fryer}, C., {et~al.} 2012, \apj, 759, 52, \dodoi{10.1088/0004-637X/759/1/52}

\bibitem[{{Dorda} {et~al.}(2016){Dorda}, {Gonz{\'a}lez-Fern{\'a}ndez}, \& {Negueruela}}]{Dorda2016}
{Dorda}, R., {Gonz{\'a}lez-Fern{\'a}ndez}, C., \& {Negueruela}, I. 2016, \aap, 595, A105, \dodoi{10.1051/0004-6361/201628422}

\bibitem[{{Dotter}(2016)}]{Dotter2016}
{Dotter}, A. 2016, \apjs, 222, 8, \dodoi{10.3847/0067-0049/222/1/8}

\bibitem[{{Dye} {et~al.}(2018){Dye}, {Lawrence}, {Read}, {Fan}, {Kerr}, {Varricatt}, {Furnell}, {Edge}, {Irwin}, {Hambly}, {Lucas}, {Almaini}, {Chambers}, {Green}, {Hewett}, {Liu}, {McGreer}, {Best}, {Zhang}, {Sutorius}, {Froebrich}, {Magnier}, {Hasinger}, {Lederer}, {Bold}, \& {Tedds}}]{ukirt_hemisphere}
{Dye}, S., {Lawrence}, A., {Read}, M.~A., {et~al.} 2018, \mnras, 473, 5113, \dodoi{10.1093/mnras/stx2622}

\bibitem[{{Fabricant} {et~al.}(2019){Fabricant}, {Fata}, {Epps}, {Gauron}, {Mueller}, {Zajac}, {Amato}, {Barberis}, {Bergner}, {Brennan}, {Brown}, {Chilingarian}, {Geary}, {Kradinov}, {McLeod}, {Smith}, \& {Woods}}]{Fabricant2019}
{Fabricant}, D., {Fata}, R., {Epps}, H., {et~al.} 2019, \pasp, 131, 075004, \dodoi{10.1088/1538-3873/ab1d78}

\bibitem[{{Ferrarese} {et~al.}(2000){Ferrarese}, {Mould}, {Kennicutt}, {Huchra}, {Ford}, {Freedman}, {Stetson}, {Madore}, {Sakai}, {Gibson}, {Graham}, {Hughes}, {Illingworth}, {Kelson}, {Macri}, {Sebo}, \& {Silbermann}}]{Ferrarese2000}
{Ferrarese}, L., {Mould}, J.~R., {Kennicutt}, Jr., R.~C., {et~al.} 2000, \apj, 529, 745, \dodoi{10.1086/308309}

\bibitem[{{Flewelling} {et~al.}(2020){Flewelling}, {Magnier}, {Chambers}, {Heasley}, {Holmberg}, {Huber}, {Sweeney}, {Waters}, {Calamida}, {Casertano}, {Chen}, {Farrow}, {Hasinger}, {Henderson}, {Long}, {Metcalfe}, {Narayan}, {Nieto-Santisteban}, {Norberg}, {Rest}, {Saglia}, {Szalay}, {Thakar}, {Tonry}, {Valenti}, {Werner}, {White}, {Denneau}, {Draper}, {Hodapp}, {Jedicke}, {Kaiser}, {Kudritzki}, {Price}, {Wainscoat}, {Chastel}, {McLean}, {Postman}, \& {Shiao}}]{Flewelling2020}
{Flewelling}, H.~A., {Magnier}, E.~A., {Chambers}, K.~C., {et~al.} 2020, \apjs, 251, 7, \dodoi{10.3847/1538-4365/abb82d}

\bibitem[{{Foreman-Mackey} {et~al.}(2013){Foreman-Mackey}, {Hogg}, {Lang}, \& {Goodman}}]{Mackey-Foreman2013}
{Foreman-Mackey}, D., {Hogg}, D.~W., {Lang}, D., \& {Goodman}, J. 2013, \pasp, 125, 306, \dodoi{10.1086/670067}

\bibitem[{{Frostig} {et~al.}(2024){Frostig}, {Burdge}, {De}, {Fur{\'e}sz}, {Haworth}, {Hinrichsen}, {Karambelkar}, {Kasliwal}, {Lourie}, {Malonis}, {Mo}, {Simcoe}, {Soto}, \& {Stein}}]{Frostig2024}
{Frostig}, D., {Burdge}, K.~B., {De}, K., {et~al.} 2024, in Society of Photo-Optical Instrumentation Engineers (SPIE) Conference Series, Vol. 13096, Ground-based and Airborne Instrumentation for Astronomy X, ed. J.~J. {Bryant}, K.~{Motohara}, \& J.~R.~D. {Vernet}, 130963J, \dodoi{10.1117/12.3019165}

\bibitem[{{Gavetti} {et~al.}(2025){Gavetti}, {Ventura}, {Dell'Agli}, {La Franca}, {Marini}, {Correnti}, \& {Tailo}}]{Gavetti2025}
{Gavetti}, C., {Ventura}, P., {Dell'Agli}, F., {et~al.} 2025, arXiv e-prints, arXiv:2504.12940.
\newblock \doarXiv{2504.12940}

\bibitem[{{Gonz{\'a}lez-Bol{\'\i}var} {et~al.}(2024){Gonz{\'a}lez-Bol{\'\i}var}, {De Marco}, {Berm{\'u}dez-Bustamante}, {Siess}, \& {Price}}]{Gonzalez-Bolivar2024}
{Gonz{\'a}lez-Bol{\'\i}var}, M., {De Marco}, O., {Berm{\'u}dez-Bustamante}, L.~C., {Siess}, L., \& {Price}, D.~J. 2024, \mnras, 527, 9145, \dodoi{10.1093/mnras/stad3748}

\bibitem[{{Herter} {et~al.}(2008){Herter}, {Henderson}, {Wilson}, {Matthews}, {Rahmer}, {Bonati}, {Muirhead}, {Adams}, {Lloyd}, {Skrutskie}, {Moon}, {Parshley}, {Nelson}, {Martinache}, \& {Gull}}]{Herter2008}
{Herter}, T.~L., {Henderson}, C.~P., {Wilson}, J.~C., {et~al.} 2008, in Society of Photo-Optical Instrumentation Engineers (SPIE) Conference Series, Vol. 7014, \procspie, 70140X, \dodoi{10.1117/12.789660}

\bibitem[{{Husser} {et~al.}(2013){Husser}, {Wende-von Berg}, {Dreizler}, {Homeier}, {Reiners}, {Barman}, \& {Hauschildt}}]{Husser2013}
{Husser}, T.~O., {Wende-von Berg}, S., {Dreizler}, S., {et~al.} 2013, \aap, 553, A6, \dodoi{10.1051/0004-6361/201219058}

\bibitem[{{Iaconi} {et~al.}(2018){Iaconi}, {De Marco}, {Passy}, \& {Staff}}]{Iaconi2018}
{Iaconi}, R., {De Marco}, O., {Passy}, J.-C., \& {Staff}, J. 2018, \mnras, 477, 2349, \dodoi{10.1093/mnras/sty794}

\bibitem[{{Iaconi} {et~al.}(2019){Iaconi}, {Maeda}, {De Marco}, {Nozawa}, \& {Reichardt}}]{Iaconi2019}
{Iaconi}, R., {Maeda}, K., {De Marco}, O., {Nozawa}, T., \& {Reichardt}, T. 2019, \mnras, 489, 3334, \dodoi{10.1093/mnras/stz2312}

\bibitem[{{Iaconi} {et~al.}(2020){Iaconi}, {Maeda}, {Nozawa}, {De Marco}, \& {Reichardt}}]{Iaconi2020}
{Iaconi}, R., {Maeda}, K., {Nozawa}, T., {De Marco}, O., \& {Reichardt}, T. 2020, \mnras, 497, 3166, \dodoi{10.1093/mnras/staa2169}

\bibitem[{{Ivanova} {et~al.}(2013){Ivanova}, {Justham}, {Chen}, {De Marco}, {Fryer}, {Gaburov}, {Ge}, {Glebbeek}, {Han}, {Li}, {Lu}, {Marsh}, {Podsiadlowski}, {Potter}, {Soker}, {Taam}, {Tauris}, {van den Heuvel}, \& {Webbink}}]{Ivanova2013araa}
{Ivanova}, N., {Justham}, S., {Chen}, X., {et~al.} 2013, \aapr, 21, 59, \dodoi{10.1007/s00159-013-0059-2}

\bibitem[{{Ivezic} \& {Elitzur}(1997)}]{Ivezic97}
{Ivezic}, Z., \& {Elitzur}, M. 1997, MNRAS, 287, 799

\bibitem[{{Ivezic} {et~al.}(1999){Ivezic}, {Nenkova}, \& {Elitzur}}]{Ivezic99}
{Ivezic}, Z., {Nenkova}, M., \& {Elitzur}, M. 1999, astro-ph/9910475

\bibitem[{{Jacoby} {et~al.}(1984){Jacoby}, {Hunter}, \& {Christian}}]{Jacoby1984}
{Jacoby}, G.~H., {Hunter}, D.~A., \& {Christian}, C.~A. 1984, \apjs, 56, 257, \dodoi{10.1086/190983}

\bibitem[{{Jencson} {et~al.}(2019){Jencson}, {Kasliwal}, {Adams}, {Bond}, {De}, {Johansson}, {Karambelkar}, {Lau}, {Tinyanont}, {Ryder}, {Cody}, {Masci}, {Bally}, {Blagorodnova}, {Castell{\'o}n}, {Fremling}, {Gehrz}, {Helou}, {Kilpatrick}, {Milne}, {Morrell}, {Perley}, {Phillips}, {Smith}, {van Dyk}, \& {Williams}}]{Jencson2019_spirits}
{Jencson}, J.~E., {Kasliwal}, M.~M., {Adams}, S.~M., {et~al.} 2019, \apj, 886, 40, \dodoi{10.3847/1538-4357/ab4a01}

\bibitem[{{Jencson} {et~al.}(2022){Jencson}, {Sand}, {Andrews}, {Smith}, {Pearson}, {Strader}, {Valenti}, {Beasor}, \& {Rothberg}}]{Jencson2022}
{Jencson}, J.~E., {Sand}, D.~J., {Andrews}, J.~E., {et~al.} 2022, \apj, 930, 81, \dodoi{10.3847/1538-4357/ac626c}

\bibitem[{{Karambelkar} {et~al.}(2020){Karambelkar}, {Kasliwal}, {Tisserand}, {De}, {Anand}, {Ashley}, {Delacroix}, {Hankins}, {Jencson}, {Lau}, {McKenna}, {Moore}, {Ofek}, {Smith}, {Soria}, {Soon}, {Tinyanont}, {Travouillon}, \& {Yao}}]{Karambelkar2020}
{Karambelkar}, V.~R., {Kasliwal}, M.~M., {Tisserand}, P., {et~al.} 2020, arXiv e-prints, arXiv:2012.11629.
\newblock \doarXiv{2012.11629}

\bibitem[{{Karambelkar} {et~al.}(2023){Karambelkar}, {Kasliwal}, {Blagorodnova}, {Sollerman}, {Aloisi}, {Anand}, {Andreoni}, {Brink}, {Bruch}, {Cook}, {Das}, {De}, {Drake}, {Filippenko}, {Fremling}, {Helou}, {Ho}, {Jencson}, {Jones}, {Laher}, {Masci}, {Patra}, {Purdum}, {Reedy}, {Sit}, {Sharma}, {Tzanidakis}, {van der Walt}, {Yao}, \& {Zhang}}]{Karambelkar2023}
{Karambelkar}, V.~R., {Kasliwal}, M.~M., {Blagorodnova}, N., {et~al.} 2023, \apj, 948, 137, \dodoi{10.3847/1538-4357/acc2b9}

\bibitem[{{Kasliwal} {et~al.}(2011){Kasliwal}, {Kulkarni}, {Arcavi}, {Quimby}, {Ofek}, {Nugent}, {Jacobsen}, {Gal-Yam}, {Green}, {Yaron}, {Fox}, {Howell}, {Cenko}, {Kleiser}, {Bloom}, {Miller}, {Li}, {Filippenko}, {Starr}, {Poznanski}, {Law}, {Helou}, {Frail}, {Neill}, {Forster}, {Martin}, {Tendulkar}, {Gehrels}, {Kennea}, {Sullivan}, {Bildsten}, {Dekany}, {Rahmer}, {Hale}, {Smith}, {Zolkower}, {Velur}, {Walters}, {Henning}, {Bui}, {McKenna}, \& {Blake}}]{Kasliwal2011}
{Kasliwal}, M.~M., {Kulkarni}, S.~R., {Arcavi}, I., {et~al.} 2011, \apj, 730, 134, \dodoi{10.1088/0004-637X/730/2/134}

\bibitem[{{Kasliwal} {et~al.}(2017){Kasliwal}, {Bally}, {Masci}, {Cody}, {Bond}, {Jencson}, {Tinyanont}, {Cao}, {Contreras}, {Dykhoff}, {Amodeo}, {Armus}, {Boyer}, {Cantiello}, {Carlon}, {Cass}, {Cook}, {Corgan}, {Faella}, {Fox}, {Green}, {Gehrz}, {Helou}, {Hsiao}, {Johansson}, {Khan}, {Lau}, {Langer}, {Levesque}, {Milne}, {Mohamed}, {Morrell}, {Monson}, {Moore}, {Ofek}, {O' Sullivan}, {Parthasarathy}, {Perez}, {Perley}, {Phillips}, {Prince}, {Shenoy}, {Smith}, {Surace}, {Van Dyk}, {Whitelock}, \& {Williams}}]{Kasliwal2017ApJ}
{Kasliwal}, M.~M., {Bally}, J., {Masci}, F., {et~al.} 2017, \apj, 839, 88, \dodoi{10.3847/1538-4357/aa6978}

\bibitem[{{Kasliwal} {et~al.}(2025){Kasliwal}, {Earley}, {Smith}, {Guillot}, {Travouillon}, {Fucik}, {Abe}, {Greffe}, {Agabi}, {Ashley}, {Triaud}, {Tinyanont}, {Antier}, {Bendjoya}, {Bhattarai}, {Bertz}, {Brugger}, {Burdanov}, {Caiazzo}, {Carry}, {Casagrande}, {Cooke}, {De}, {Dekany}, {Deloupy}, {Dornic}, {Fahey}, {Figer}, {Freeman}, {Frostig}, {G{\"u}nther}, {Hale}, {Bland-Hawthorn}, {Illuminati}, {Jencson}, {Karambelkar}, {Key}, {Lau}, {Li}, {Lubin}, {Nash}, {Neill}, {Pahuja}, {Pian}, {de Ugarte Postigo}, {Roberts}, {Rodriguez}, {Rose}, {Ruiter}, {Schmider}, {Simcoe}, {Stein}, {Suarez}, {Taylor}, {Weber}, {Wen}, {de Wit}, {Zarzaca}, \& {Zimmer}}]{Kasliwal2025_cryoscope}
{Kasliwal}, M.~M., {Earley}, N., {Smith}, R., {et~al.} 2025, arXiv e-prints, arXiv:2502.06950, \dodoi{10.48550/arXiv.2502.06950}

\bibitem[{{Kato}(2003)}]{Kato2003}
{Kato}, T. 2003, \aap, 399, 695, \dodoi{10.1051/0004-6361:20021808}

\bibitem[{{Khan}(2017)}]{Khan2017}
{Khan}, R. 2017, \apjs, 228, 5, \dodoi{10.3847/1538-4365/228/1/5}

\bibitem[{{Kingma} \& {Ba}(2014)}]{Kingma2014}
{Kingma}, D.~P., \& {Ba}, J. 2014, arXiv e-prints, arXiv:1412.6980, \dodoi{10.48550/arXiv.1412.6980}

\bibitem[{{Klencki} {et~al.}(2021){Klencki}, {Nelemans}, {Istrate}, \& {Chruslinska}}]{Klencki2021}
{Klencki}, J., {Nelemans}, G., {Istrate}, A.~G., \& {Chruslinska}, M. 2021, \aap, 645, A54, \dodoi{10.1051/0004-6361/202038707}

\bibitem[{{Kochanek} {et~al.}(2014){Kochanek}, {Adams}, \& {Belczynski}}]{Kochanek14_mergers}
{Kochanek}, C.~S., {Adams}, S.~M., \& {Belczynski}, K. 2014, \mnras, 443, 1319, \dodoi{10.1093/mnras/stu1226}

\bibitem[{{Kondo} {et~al.}(2023){Kondo}, {Sumi}, {Koshimoto}, {Rattenbury}, {Suzuki}, \& {Bennett}}]{Konndo2023}
{Kondo}, I., {Sumi}, T., {Koshimoto}, N., {et~al.} 2023, \aj, 165, 254, \dodoi{10.3847/1538-3881/acccf9}

\bibitem[{{Kulkarni} {et~al.}(2007){Kulkarni}, {Ofek}, {Rau}, {Cenko}, {Soderberg}, {Fox}, {Gal-Yam}, {Capak}, {Moon}, {Li}, {Filippenko}, {Egami}, {Kartaltepe}, \& {Sanders}}]{Kulkarni07}
{Kulkarni}, S.~R., {Ofek}, E.~O., {Rau}, A., {et~al.} 2007, \nat, 447, 458, \dodoi{10.1038/nature05822}

\bibitem[{{Ku{\v{c}}inskas} {et~al.}(2006){Ku{\v{c}}inskas}, {Hauschildt}, {Brott}, {Vansevi{\v{c}}ius}, {Lindegren}, {Tanab{\'e}}, \& {Allard}}]{Kucinskas2006}
{Ku{\v{c}}inskas}, A., {Hauschildt}, P.~H., {Brott}, I., {et~al.} 2006, \aap, 452, 1021, \dodoi{10.1051/0004-6361:20054431}

\bibitem[{{Ku{\v{c}}inskas} {et~al.}(2005){Ku{\v{c}}inskas}, {Hauschildt}, {Ludwig}, {Brott}, {Vansevi{\v{c}}ius}, {Lindegren}, {Tanab{\'e}}, \& {Allard}}]{Kucinskas2005}
{Ku{\v{c}}inskas}, A., {Hauschildt}, P.~H., {Ludwig}, H.~G., {et~al.} 2005, \aap, 442, 281, \dodoi{10.1051/0004-6361:20053028}

\bibitem[{{Laher} {et~al.}(2014){Laher}, {Surace}, {Grillmair}, {Ofek}, {Levitan}, {Sesar}, {van Eyken}, {Law}, {Helou}, {Hamam}, {Masci}, {Mattingly}, {Jackson}, {Hacopeans}, {Mi}, {Groom}, {Teplitz}, {Desai}, {Hale}, {Smith}, {Walters}, {Quimby}, {Kasliwal}, {Horesh}, {Bellm}, {Barlow}, {Waszczak}, {Prince}, \& {Kulkarni}}]{Laher14}
{Laher}, R.~R., {Surace}, J., {Grillmair}, C.~J., {et~al.} 2014, \pasp, 126, 674, \dodoi{10.1086/677351}

\bibitem[{{Lau} {et~al.}(2025){Lau}, {Jencson}, {Salyk}, {De}, {Fox}, {Hankins}, {Kasliwal}, {Keyes}, {Macleod}, {Ressler}, \& {Rose}}]{Lau2025}
{Lau}, R.~M., {Jencson}, J.~E., {Salyk}, C., {et~al.} 2025, \apj, 983, 87, \dodoi{10.3847/1538-4357/adb429}

\bibitem[{{Law} {et~al.}(2009){Law}, {Kulkarni}, {Dekany}, {Ofek}, {Quimby}, {Nugent}, {Surace}, {Grillmair}, {Bloom}, {Kasliwal}, {Bildsten}, {Brown}, {Cenko}, {Ciardi}, {Croner}, {Djorgovski}, {van Eyken}, {Filippenko}, {Fox}, {Gal-Yam}, {Hale}, {Hamam}, {Helou}, {Henning}, {Howell}, {Jacobsen}, {Laher}, {Mattingly}, {McKenna}, {Pickles}, {Poznanski}, {Rahmer}, {Rau}, {Rosing}, {Shara}, {Smith}, {Starr}, {Sullivan}, {Velur}, {Walters}, \& {Zolkower}}]{Law09}
{Law}, N.~M., {Kulkarni}, S.~R., {Dekany}, R.~G., {et~al.} 2009, \pasp, 121, 1395, \dodoi{10.1086/648598}

\bibitem[{{Loebman} {et~al.}(2015){Loebman}, {Wisniewski}, {Schmidt}, {Kowalski}, {Barry}, {Bjorkman}, {Hammel}, {Hawley}, {Hebb}, {Kasliwal}, {Lynch}, {Russell}, {Sitko}, \& {Szkody}}]{Loebman15}
{Loebman}, S.~R., {Wisniewski}, J.~P., {Schmidt}, S.~J., {et~al.} 2015, \aj, 149, 17, \dodoi{10.1088/0004-6256/149/1/17}

\bibitem[{{Lourie} {et~al.}(2020){Lourie}, {Baker}, {Burruss}, {Egan}, {F{\.z}r{\'e}sz}, {Frostig}, {Garcia-Zych}, {Ganciu}, {Haworth}, {Hinrichsen}, {Kasliwal}, {Karambelkar}, {Malonis}, {Simcoe}, \& {Zolkower}}]{Lourie2020}
{Lourie}, N.~P., {Baker}, J.~W., {Burruss}, R.~S., {et~al.} 2020, in Society of Photo-Optical Instrumentation Engineers (SPIE) Conference Series, Vol. 11447, Society of Photo-Optical Instrumentation Engineers (SPIE) Conference Series, 114479K, \dodoi{10.1117/12.2561210}

\bibitem[{{Lynch} {et~al.}(2004){Lynch}, {Rudy}, {Russell}, {Mazuk}, {Venturini}, {Dimpfl}, {Bernstein}, {Sitko}, {Fajardo-Acosta}, {Tokunaga}, {Knacke}, {Puetter}, \& {Perry}}]{Lynch04}
{Lynch}, D.~K., {Rudy}, R.~J., {Russell}, R.~W., {et~al.} 2004, \apj, 607, 460, \dodoi{10.1086/382667}

\bibitem[{{Lynch} {et~al.}(2007){Lynch}, {Rudy}, {Russell}, {Mazuk}, {Venturini}, {Bernstein}, {Puetter}, {Perry}, \& {Skinner}}]{Lynch07}
{Lynch}, D.~K., {Rudy}, R.~J., {Russell}, R.~W., {et~al.} 2007, in Astronomical Society of the Pacific Conference Series, Vol. 363, The Nature of V838 Mon and its Light Echo, ed. R.~L.~M. {Corradi} \& U.~{Munari}, 39

\bibitem[{Lü {et~al.}(2013)Lü, Zhu, \& Podsiadlowski}]{Lu2013}
Lü, G., Zhu, C., \& Podsiadlowski, P. 2013, The Astrophysical Journal, 768, 193, \dodoi{10.1088/0004-637X/768/2/193}

\bibitem[{{MacLeod} {et~al.}(2018){MacLeod}, {Cantiello}, \& {Soares-Furtado}}]{MacLeod2018}
{MacLeod}, M., {Cantiello}, M., \& {Soares-Furtado}, M. 2018, \apjl, 853, L1, \dodoi{10.3847/2041-8213/aaa5fa}

\bibitem[{{MacLeod} {et~al.}(2022){MacLeod}, {De}, \& {Loeb}}]{MacLeod2022}
{MacLeod}, M., {De}, K., \& {Loeb}, A. 2022, \apj, 937, 96, \dodoi{10.3847/1538-4357/ac8c31}

\bibitem[{{MacLeod} {et~al.}(2017){MacLeod}, {Macias}, {Ramirez-Ruiz}, {Grindlay}, {Batta}, \& {Montes}}]{MacLeod17}
{MacLeod}, M., {Macias}, P., {Ramirez-Ruiz}, E., {et~al.} 2017, \apj, 835, 282, \dodoi{10.3847/1538-4357/835/2/282}

\bibitem[{{Magnier} {et~al.}(2013){Magnier}, {Schlafly}, {Finkbeiner}, {Juric}, {Tonry}, {Burgett}, {Chambers}, {Flewelling}, {Kaiser}, {Kudritzki}, {Morgan}, {Price}, {Sweeney}, \& {Stubbs}}]{Magnier13}
{Magnier}, E.~A., {Schlafly}, E., {Finkbeiner}, D., {et~al.} 2013, \apjs, 205, 20, \dodoi{10.1088/0067-0049/205/2/20}

\bibitem[{{Mainzer} {et~al.}(2014){Mainzer}, {Bauer}, {Cutri}, {Grav}, {Masiero}, {Beck}, {Clarkson}, {Conrow}, {Dailey}, {Eisenhardt}, {Fabinsky}, {Fajardo-Acosta}, {Fowler}, {Gelino}, {Grillmair}, {Heinrichsen}, {Kendall}, {Kirkpatrick}, {Liu}, {Masci}, {McCallon}, {Nugent}, {Papin}, {Rice}, {Royer}, {Ryan}, {Sevilla}, {Sonnett}, {Stevenson}, {Thompson}, {Wheelock}, {Wiemer}, {Wittman}, {Wright}, \& {Yan}}]{Mainzer2014ApJ}
{Mainzer}, A., {Bauer}, J., {Cutri}, R.~M., {et~al.} 2014, \apj, 792, 30, \dodoi{10.1088/0004-637X/792/1/30}

\bibitem[{{Mandigo-Stoba} {et~al.}(2022){Mandigo-Stoba}, {Fremling}, \& {Kasliwal}}]{MandigoStoba2021}
{Mandigo-Stoba}, M.~S., {Fremling}, C., \& {Kasliwal}, M. 2022, The Journal of Open Source Software, 7, 3612, \dodoi{10.21105/joss.03612}

\bibitem[{{Marchant} {et~al.}(2021){Marchant}, {Pappas}, {Gallegos-Garcia}, {Berry}, {Taam}, {Kalogera}, \& {Podsiadlowski}}]{Marchant2021}
{Marchant}, P., {Pappas}, K. M.~W., {Gallegos-Garcia}, M., {et~al.} 2021, \aap, 650, A107, \dodoi{10.1051/0004-6361/202039992}

\bibitem[{{Marigo} {et~al.}(2013){Marigo}, {Bressan}, {Nanni}, {Girardi}, \& {Pumo}}]{Marigo2013}
{Marigo}, P., {Bressan}, A., {Nanni}, A., {Girardi}, L., \& {Pumo}, M.~L. 2013, \mnras, 434, 488, \dodoi{10.1093/mnras/stt1034}

\bibitem[{{Marigo} \& {Girardi}(2007)}]{Marigo2007}
{Marigo}, P., \& {Girardi}, L. 2007, \aap, 469, 239, \dodoi{10.1051/0004-6361:20066772}

\bibitem[{{Marigo} {et~al.}(2008){Marigo}, {Girardi}, {Bressan}, {Groenewegen}, {Silva}, \& {Granato}}]{Marigo08}
{Marigo}, P., {Girardi}, L., {Bressan}, A., {et~al.} 2008, \aap, 482, 883, \dodoi{10.1051/0004-6361:20078467}

\bibitem[{{Masci} {et~al.}(2019){Masci}, {Laher}, {Rusholme}, {Shupe}, {Groom}, {Surace}, {Jackson}, {Monkewitz}, {Beck}, {Flynn}, {Terek}, {Landry}, {Hacopians}, {Desai}, {Howell}, {Brooke}, {Imel}, {Wachter}, {Ye}, {Lin}, {Cenko}, {Cunningham}, {Rebbapragada}, {Bue}, {Miller}, {Mahabal}, {Bellm}, {Patterson}, {Juri{\'c}}, {Golkhou}, {Ofek}, {Walters}, {Graham}, {Kasliwal}, {Dekany}, {Kupfer}, {Burdge}, {Cannella}, {Barlow}, {Van Sistine}, {Giomi}, {Fremling}, {Blagorodnova}, {Levitan}, {Riddle}, {Smith}, {Helou}, {Prince}, \& {Kulkarni}}]{Masci2019}
{Masci}, F.~J., {Laher}, R.~R., {Rusholme}, B., {et~al.} 2019, \pasp, 131, 018003, \dodoi{10.1088/1538-3873/aae8ac}

\bibitem[{{Meisner} {et~al.}(2019){Meisner}, {Lang}, {Schlafly}, \& {Schlegel}}]{Meisner2019}
{Meisner}, A.~M., {Lang}, D., {Schlafly}, E.~F., \& {Schlegel}, D.~J. 2019, \pasp, 131, 124504, \dodoi{10.1088/1538-3873/ab3df4}

\bibitem[{{Metzger} \& {Pejcha}(2017)}]{Metzger2017}
{Metzger}, B.~D., \& {Pejcha}, O. 2017, \mnras, 471, 3200, \dodoi{10.1093/mnras/stx1768}

\bibitem[{{Munari} {et~al.}(2002){Munari}, {Henden}, {Kiyota}, {Laney}, {Marang}, {Zwitter}, {Corradi}, {Desidera}, {Marrese}, {Giro}, {Boschi}, \& {Schwartz}}]{Munari2002AA}
{Munari}, U., {Henden}, A., {Kiyota}, S., {et~al.} 2002, \aap, 389, L51, \dodoi{10.1051/0004-6361:20020715}

\bibitem[{{Neugent} {et~al.}(2020){Neugent}, {Massey}, {Georgy}, {Drout}, {Mommert}, {Levesque}, {Meynet}, \& {Ekstr{\"o}m}}]{Neugent2020}
{Neugent}, K.~F., {Massey}, P., {Georgy}, C., {et~al.} 2020, \apj, 889, 44, \dodoi{10.3847/1538-4357/ab5ba0}

\bibitem[{{Ofek} {et~al.}(2012){Ofek}, {Laher}, {Law}, {Surace}, {Levitan}, {Sesar}, {Horesh}, {Poznanski}, {van Eyken}, {Kulkarni}, {Nugent}, {Zolkower}, {Walters}, {Sullivan}, {Ag{\"u}eros}, {Bildsten}, {Bloom}, {Cenko}, {Gal-Yam}, {Grillmair}, {Helou}, {Kasliwal}, \& {Quimby}}]{Ofek12}
{Ofek}, E.~O., {Laher}, R., {Law}, N., {et~al.} 2012, \pasp, 124, 62, \dodoi{10.1086/664065}

\bibitem[{{Oke} \& {Gunn}(1982)}]{Oke82}
{Oke}, J.~B., \& {Gunn}, J.~E. 1982, \pasp, 94, 586, \dodoi{10.1086/131027}

\bibitem[{{Oke} {et~al.}(1995){Oke}, {Cohen}, {Carr}, {Cromer}, {Dingizian}, {Harris}, {Labrecque}, {Lucinio}, {Schaal}, {Epps}, \& {Miller}}]{Oke95}
{Oke}, J.~B., {Cohen}, J.~G., {Carr}, M., {et~al.} 1995, \pasp, 107, 375, \dodoi{10.1086/133562}

\bibitem[{{Paczynski}(1976)}]{Paczynski1976IAUS}
{Paczynski}, B. 1976, in IAU Symposium, Vol.~73, Structure and Evolution of Close Binary Systems, ed. P.~{Eggleton}, S.~{Mitton}, \& J.~{Whelan}, 75

\bibitem[{{Pastorello} {et~al.}(2019){Pastorello}, {Mason}, {Taubenberger}, {Fraser}, {Cortini}, {Tomasella}, {Botticella}, {Elias-Rosa}, {Kotak}, {Smartt}, {Benetti}, {Cappellaro}, {Turatto}, {Tartaglia}, {Djorgovski}, {Drake}, {Berton}, {Briganti}, {Brimacombe}, {Bufano}, {Cai}, {Chen}, {Christensen}, {Ciabattari}, {Congiu}, {Dimai}, {Inserra}, {Kankare}, {Magill}, {Maguire}, {Martinelli}, {Morales-Garoffolo}, {Ochner}, {Pignata}, {Reguitti}, {Sollerman}, {Spiro}, {Terreran}, \& {Wright}}]{Pastorello2019a}
{Pastorello}, A., {Mason}, E., {Taubenberger}, S., {et~al.} 2019, \aap, 630, A75, \dodoi{10.1051/0004-6361/201935999}

\bibitem[{{Pastorello} {et~al.}(2021{\natexlab{a}}){Pastorello}, {Fraser}, {Valerin}, {Reguitti}, {Itagaki}, {Ochner}, {Williams}, {Jones}, {Munday}, {Smartt}, {Smith}, {Srivastav}, {Elias-Rosa}, {Kankare}, {Karamehmetoglu}, {Lundqvist}, {Mazzali}, {Munari}, {Stritzinger}, {Tomasella}, {Anderson}, {Chambers}, \& {Rest}}]{Pastorello2021a}
{Pastorello}, A., {Fraser}, M., {Valerin}, G., {et~al.} 2021{\natexlab{a}}, \aap, 646, A119, \dodoi{10.1051/0004-6361/202039952}

\bibitem[{{Pastorello} {et~al.}(2021{\natexlab{b}}){Pastorello}, {Valerin}, {Fraser}, {Elias-Rosa}, {Valenti}, {Reguitti}, {Mazzali}, {Amaro}, {Andrews}, {Dong}, {Jencson}, {Lundquist}, {Reichart}, {Sand}, {Wyatt}, {Smartt}, {Smith}, {Srivastav}, {Cai}, {Cappellaro}, {Holmbo}, {Fiore}, {Jones}, {Kankare}, {Karamehmetoglu}, {Lundqvist}, {Morales-Garoffolo}, {Reynolds}, {Stritzinger}, {Williams}, {Chambers}, {de Boer}, {Huber}, {Rest}, \& {Wainscoat}}]{Pastorello2021b}
{Pastorello}, A., {Valerin}, G., {Fraser}, M., {et~al.} 2021{\natexlab{b}}, \aap, 647, A93, \dodoi{10.1051/0004-6361/202039953}

\bibitem[{{Pastorello} {et~al.}(2022){Pastorello}, {Valerin}, {Fraser}, {Reguitti}, {Elias-Rosa}, {Filippenko}, {Rojas-Bravo}, {Tartaglia}, {Reynolds}, {Valenti}, {Andrews}, {Ashall}, {Bostroem}, {Brink}, {Burke}, {Cai}, {Cappellaro}, {Coulter}, {Dastidar}, {Davis}, {Dimitriadis}, {Fiore}, {Foley}, {Fugazza}, {Galbany}, {Gangopadhyay}, {Geier}, {Gutierrez}, {Haislip}, {Hiramatsu}, {Holmbo}, {Howell}, {Hsiao}, {Hung}, {Jha}, {Kankare}, {Karamehmetoglu}, {Kilpatrick}, {Kotak}, {Kouprianov}, {Kravtsov}, {Kumar}, {Li}, {Lundquist}, {Lundqvist}, {Matilainen}, {Mazzali}, {McCully}, {Misra}, {Morales-Garoffolo}, {Moran}, {Morrell}, {Newsome}, {Padilla Gonzalez}, {Pan}, {Pellegrino}, {Phillips}, {Pignata}, {Piro}, {Reichart}, {Rest}, {Salmaso}, {Sand}, {Siebert}, {Smartt}, {Smith}, {Srivastav}, {Stritzinger}, {Taggart}, {Tinyanont}, {Yan}, {Wang}, {Wang}, {Williams}, {Wyatt}, {Zhang}, {de Boer}, {Chambers}, {Gao}, \& {Magnier}}]{Pastorello2022}
---. 2022, arXiv e-prints, arXiv:2208.02782.
\newblock \doarXiv{2208.02782}

\bibitem[{Pedregosa {et~al.}(2011)Pedregosa, Varoquaux, Gramfort, Michel, Thirion, Grisel, Blondel, Prettenhofer, Weiss, Dubourg, Vanderplas, Passos, Cournapeau, Brucher, Perrot, \& Duchesnay}]{scikit-learn}
Pedregosa, F., Varoquaux, G., Gramfort, A., {et~al.} 2011, Journal of Machine Learning Research, 12, 2825

\bibitem[{{Pejcha}(2014)}]{Pejcha14}
{Pejcha}, O. 2014, \apj, 788, 22, \dodoi{10.1088/0004-637X/788/1/22}

\bibitem[{{Pejcha} {et~al.}(2016){Pejcha}, {Metzger}, \& {Tomida}}]{Pejcha16b}
{Pejcha}, O., {Metzger}, B.~D., \& {Tomida}, K. 2016, \mnras, 455, 4351, \dodoi{10.1093/mnras/stv2592}

\bibitem[{{Perley} {et~al.}(2020){Perley}, {Fremling}, {Sollerman}, {Miller}, {Dahiwale}, {Sharma}, {Bellm}, {Biswas}, {Brink}, {Bruch}, {De}, {Dekany}, {Drake}, {Duev}, {Filippenko}, {Gal-Yam}, {Goobar}, {Graham}, {Graham}, {Ho}, {Irani}, {Kasliwal}, {Kim}, {Kulkarni}, {Mahabal}, {Masci}, {Modak}, {Neill}, {Nordin}, {Riddle}, {Soumagnac}, {Strotjohann}, {Schulze}, {Taggart}, {Tzanidakis}, {Walters}, \& {Yan}}]{Perley2020}
{Perley}, D.~A., {Fremling}, C., {Sollerman}, J., {et~al.} 2020, \apj, 904, 35, \dodoi{10.3847/1538-4357/abbd98}

\bibitem[{{Petz} \& {Kochanek}(2025)}]{Sydney2025}
{Petz}, S., \& {Kochanek}, C.~S. 2025, arXiv e-prints, arXiv:2501.14058, \dodoi{10.48550/arXiv.2501.14058}

\bibitem[{{Phillips} \& {Hartmann}(1978)}]{Phillips1978}
{Phillips}, M.~J., \& {Hartmann}, L. 1978, \apj, 224, 182, \dodoi{10.1086/156363}

\bibitem[{{Postnov} \& {Yungelson}(2014)}]{Postnov2014}
{Postnov}, K.~A., \& {Yungelson}, L.~R. 2014, Living Reviews in Relativity, 17, 3, \dodoi{10.12942/lrr-2014-3}

\bibitem[{{Rayner} {et~al.}(2009){Rayner}, {Cushing}, \& {Vacca}}]{Rayner2009}
{Rayner}, J.~T., {Cushing}, M.~C., \& {Vacca}, W.~D. 2009, \apjs, 185, 289, \dodoi{10.1088/0067-0049/185/2/289}

\bibitem[{{Ricker} \& {Taam}(2012)}]{Ricker2012}
{Ricker}, P.~M., \& {Taam}, R.~E. 2012, \apj, 746, 74, \dodoi{10.1088/0004-637X/746/1/74}

\bibitem[{{Riebel} {et~al.}(2015){Riebel}, {Boyer}, {Srinivasan}, {Whitelock}, {Meixner}, {Babler}, {Feast}, {Groenewegen}, {Ita}, {Meade}, {Shiao}, \& {Whitney}}]{Riebel2015}
{Riebel}, D., {Boyer}, M.~L., {Srinivasan}, S., {et~al.} 2015, \apj, 807, 1, \dodoi{10.1088/0004-637X/807/1/1}

\bibitem[{{Rigliaco} {et~al.}(2012){Rigliaco}, {Natta}, {Testi}, {Randich}, {Alcal{\`a}}, {Covino}, \& {Stelzer}}]{Rigliaco2012}
{Rigliaco}, E., {Natta}, A., {Testi}, L., {et~al.} 2012, \aap, 548, A56, \dodoi{10.1051/0004-6361/201219832}

\bibitem[{{Rowan} {et~al.}(2021){Rowan}, {Stanek}, {Way}, {Kochanek}, {Jayasinghe}, {Thompson}, {Barker}, {Hambsch}, {Bohlsen}, {Kafka}, {Shappee}, {Holoien}, \& {Prieto}}]{Rowan2021}
{Rowan}, D.~M., {Stanek}, K.~Z., {Way}, Z., {et~al.} 2021, Research Notes of the American Astronomical Society, 5, 147, \dodoi{10.3847/2515-5172/ac0c83}

\bibitem[{{Sargent} {et~al.}(2011){Sargent}, {Srinivasan}, \& {Meixner}}]{Sargent2011}
{Sargent}, B.~A., {Srinivasan}, S., \& {Meixner}, M. 2011, \apj, 728, 93, \dodoi{10.1088/0004-637X/728/2/93}

\bibitem[{{Sargent} {et~al.}(2010){Sargent}, {Srinivasan}, {Meixner}, {Kemper}, {Tielens}, {Speck}, {Matsuura}, {Bernard}, {Hony}, {Gordon}, {Indebetouw}, {Marengo}, {Sloan}, \& {Woods}}]{Sargent2010}
{Sargent}, B.~A., {Srinivasan}, S., {Meixner}, M., {et~al.} 2010, \apj, 716, 878, \dodoi{10.1088/0004-637X/716/1/878}

\bibitem[{{Schlafly} {et~al.}(2012){Schlafly}, {Finkbeiner}, {Juri{\'c}}, {Magnier}, {Burgett}, {Chambers}, {Grav}, {Hodapp}, {Kaiser}, {Kudritzki}, {Martin}, {Morgan}, {Price}, {Rix}, {Stubbs}, {Tonry}, \& {Wainscoat}}]{Schlafly12}
{Schlafly}, E.~F., {Finkbeiner}, D.~P., {Juri{\'c}}, M., {et~al.} 2012, \apj, 756, 158, \dodoi{10.1088/0004-637X/756/2/158}

\bibitem[{{Soker}(2020)}]{Soker2020}
{Soker}, N. 2020, \apj, 893, 20, \dodoi{10.3847/1538-4357/ab7dbb}

\bibitem[{{Soker} \& {Kashi}(2012)}]{Soker2012}
{Soker}, N., \& {Kashi}, A. 2012, \apj, 746, 100, \dodoi{10.1088/0004-637X/746/1/100}

\bibitem[{{Srinivasan} {et~al.}(2011){Srinivasan}, {Sargent}, \& {Meixner}}]{Srinivasan2011}
{Srinivasan}, S., {Sargent}, B.~A., \& {Meixner}, M. 2011, \aap, 532, A54, \dodoi{10.1051/0004-6361/201117033}

\bibitem[{{Steinmetz} {et~al.}(2025){Steinmetz}, {Kami{\'n}ski}, {Melis}, {Gromadzki}, {Menten}, \& {Su}}]{Steinmetz2025}
{Steinmetz}, T., {Kami{\'n}ski}, T., {Melis}, C., {et~al.} 2025, arXiv e-prints, arXiv:2502.18365, \dodoi{10.48550/arXiv.2502.18365}

\bibitem[{{Teixeira} {et~al.}(2018){Teixeira}, {Kumar}, {Smith}, {Lucas}, {Morris}, {Borissova}, {Monteiro}, {Caratti o Garatti}, {Contreras Pe{\~n}a}, {Froebrich}, \& {Gameiro}}]{Teixeira2018}
{Teixeira}, G.~D.~C., {Kumar}, M.~S.~N., {Smith}, L., {et~al.} 2018, \aap, 619, A41, \dodoi{10.1051/0004-6361/201833667}

\bibitem[{{Torres} \& {Sakano}(2022)}]{Torres2022}
{Torres}, G., \& {Sakano}, K. 2022, \mnras, 516, 2514, \dodoi{10.1093/mnras/stac2322}

\bibitem[{{Tranin} {et~al.}(2024){Tranin}, {Blagorodnova}, {Karambelkar}, {Groot}, {Bloemen}, {Vreeswijk}, {Pieterse}, \& {van Roestel}}]{Tranin2024}
{Tranin}, H., {Blagorodnova}, N., {Karambelkar}, V., {et~al.} 2024, arXiv e-prints, arXiv:2409.11347, \dodoi{10.48550/arXiv.2409.11347}

\bibitem[{{Tylenda}(2005)}]{Tylenda05a}
{Tylenda}, R. 2005, \aap, 436, 1009, \dodoi{10.1051/0004-6361:20052800}

\bibitem[{{Tylenda} {et~al.}(2011{\natexlab{a}}){Tylenda}, {Hajduk}, {Kami{\'n}ski}, {Udalski}, {Soszy{\'n}ski}, {Szyma{\'n}ski}, {Kubiak}, {Pietrzy{\'n}ski}, {Poleski}, {Wyrzykowski}, \& {Ulaczyk}}]{Tylenda11}
{Tylenda}, R., {Hajduk}, M., {Kami{\'n}ski}, T., {et~al.} 2011{\natexlab{a}}, \aap, 528, A114, \dodoi{10.1051/0004-6361/201016221}

\bibitem[{{Tylenda} {et~al.}(2011{\natexlab{b}}){Tylenda}, {Hajduk}, {Kami{\'n}ski}, {Udalski}, {Soszy{\'n}ski}, {Szyma{\'n}ski}, {Kubiak}, {Pietrzy{\'n}ski}, {Poleski}, {Wyrzykowski}, \& {Ulaczyk}}]{Tylenda2011}
---. 2011{\natexlab{b}}, \aap, 528, A114, \dodoi{10.1051/0004-6361/201016221}

\bibitem[{{Tylenda} {et~al.}(2013){Tylenda}, {Kami{\'n}ski}, {Udalski}, {Soszy{\'n}ski}, {Poleski}, {Szyma{\'n}ski}, {Kubiak}, {Pietrzy{\'n}ski}, {Koz{\l}owski}, {Pietrukowicz}, {Ulaczyk}, \& {Wyrzykowski}}]{Tylenda13}
{Tylenda}, R., {Kami{\'n}ski}, T., {Udalski}, A., {et~al.} 2013, \aap, 555, A16, \dodoi{10.1051/0004-6361/201321647}

\bibitem[{{Tzanidakis} {et~al.}(2023){Tzanidakis}, {Davenport}, {Bellm}, \& {Wang}}]{Tzanidakis2023}
{Tzanidakis}, A., {Davenport}, J. R.~A., {Bellm}, E.~C., \& {Wang}, Y. 2023, \apj, 955, 69, \dodoi{10.3847/1538-4357/aceda7}

\bibitem[{{Ueta} \& {Meixner}(2003)}]{Ueta2003}
{Ueta}, T., \& {Meixner}, M. 2003, \apj, 586, 1338, \dodoi{10.1086/367818}

\bibitem[{{Vacca} {et~al.}(2003){Vacca}, {Cushing}, \& {Rayner}}]{Vacca2003}
{Vacca}, W.~D., {Cushing}, M.~C., \& {Rayner}, J.~T. 2003, \pasp, 115, 389, \dodoi{10.1086/346193}

\bibitem[{Vigna-Gómez {et~al.}(2020)Vigna-Gómez, MacLeod, Neijssel, Broekgaarden, Justham, Howitt, de~Mink, Vinciguerra, \& Mandel}]{VignaGomez2020}
Vigna-Gómez, A., MacLeod, M., Neijssel, C.~J., {et~al.} 2020, Publications of the Astronomical Society of Australia, 37, e038, \dodoi{10.1017/pasa.2020.31}

\bibitem[{{Webbink}(1984)}]{Webbink84}
{Webbink}, R.~F. 1984, \apj, 277, 355, \dodoi{10.1086/161701}

\bibitem[{{Whitelock} {et~al.}(2003){Whitelock}, {Feast}, {van Loon}, \& {Zijlstra}}]{Whitelock2003}
{Whitelock}, P.~A., {Feast}, M.~W., {van Loon}, J.~T., \& {Zijlstra}, A.~A. 2003, \mnras, 342, 86, \dodoi{10.1046/j.1365-8711.2003.06514.x}

\bibitem[{{Wilson} {et~al.}(2004){Wilson}, {Henderson}, {Herter}, {Matthews}, {Skrutskie}, {Adams}, {Moon}, {Smith}, {Gautier}, {Ressler}, {Soifer}, {Lin}, {Howard}, {LaMarr}, {Stolberg}, \& {Zink}}]{Wilson2004}
{Wilson}, J.~C., {Henderson}, C.~P., {Herter}, T.~L., {et~al.} 2004, in Society of Photo-Optical Instrumentation Engineers (SPIE) Conference Series, Vol. 5492, Ground-based Instrumentation for Astronomy, ed. A.~F.~M. {Moorwood} \& M.~{Iye}, 1295--1305, \dodoi{10.1117/12.550925}

\end{thebibliography}
\end{document}